%% file: p1.tex
\documentclass[12pt]{article}
\usepackage{amssymb}
\usepackage{amsmath}
\usepackage{epsf}
\usepackage{epsfig}
\usepackage{latexsym}

\usepackage{psfrag}

    \newcommand{\Reals}{\it I\kern-.4emR}
    
    \newcommand{\Notin}{/\kern-.6em\hbox{$\in$}}
    \newcommand{\Notequiv}{/\kern-.6em\hbox{$\equiv$}}
    
    \newcommand{\Ceals}{\it I\kern-.65emC}
    \newcommand{\MM}{\it I\kern-.4emM}
    \newcommand{\NN}{\it I\kern-.4emN}
    \newcommand{\yy}{\it Y\kern-.8emY}
    \newcommand{\zz}{\makebox[.80em]{\it Z\kern-.46emZ}}
    \newcommand{\tzz}{\makebox[.80em]{\scriptsize\it Z\kern-.46emZ}}
    \newtheorem{theorem}{Theorem}[section]
    \newtheorem{lemma}[theorem]{Lemma}
    \newtheorem{fact}[theorem]{Fact}
    
    \newtheorem{corollary}[theorem]{Corollary}
    
    \newtheorem{observation}[theorem]{Observation}
    \newtheorem{remark}[theorem]{Remark}
    \newtheorem{conjecture}[theorem]{Conjecture}

\newcommand{\p}{^{\prime}}

\title{Characterizing graphs with convex and connected Cayley configuration spaces}
\author{Meera Sitharam$^{*}$ ~~
Heping Gao
\footnote{
University of Florida; 
supported in part by NSF Grants
 EIA 02-18435, CCF 04-04116, and a research gift from SolidWorks.}
}

\begin{document}

\maketitle
\thispagestyle{empty}

\begin{abstract}
We define and study exact, efficient representations of realization spaces
{\em Euclidean Distance Constraint Systems (EDCS)}, which include
Linkages and Frameworks. These are graphs with distance
assignments on the edges (frameworks) or graphs with
distance interval  assignments on the edges.
Each representation corresponds to a choice of non-edge (squared) distances  
or Cayley parameters. The set of realizable distance assignments
to the chosen parameters 
yields a parametrized {\em Cayley configuration space}.
Our notion of efficiency is  based on
the {\sl convexity and connectedness} of the Cayley configuration space,
as well as   {\sl algebraic complexity  of sampling realizations}, i.e.,
sampling the Cayley configuration space and obtaining a realization
from the sample (parametrized) configuration. Significantly, we give
{\sl purely graph-theoretic, forbidden minor} characterizations
for 2D and 3D EDCS 
that capture (i) the class of graphs
that always admit efficient Cayley configuration spaces
and (ii) the possible choices of representation parameters that yield efficient
Cayley configuration spaces for a given graph.
We show that the easy direction of the 3D characterization extends to 
arbitrary dimension $d$ and is related to the concept of $d$-realizability
of graphs.
Our results automatically yield efficient algorithms for
obtaining exact descriptions of the Cayley configuration spaces and for 
sampling realizations, without missing extreme or boundary realizations.
In addition, our results are {\sl tight:} we show counterexamples
to obvious extensions.

This is the first step
in a systematic and graded program
of combinatorial characterizations of efficient Cayley configuration spaces.
We discuss
several future theoretical and applied
research directions.

In particular, the results presented here are the first to
completely characterize
EDCS that have connected, convex and efficient Cayley configuration spaces,
based on precise and formal measures of efficiency.
It should be noted that our results {\sl do not rely on genericity}
of the EDCS.
Some of our proofs employ an unusual interplay of
(a) classical analytic results related to
positive semi-definiteness of  Euclidean distance matrices,
with (b) recent forbidden minor characterizations
and algorithms related to $d$-realizability of graphs.
We further introduce a novel type of restricted edge contraction
or reduction to a graph minor, a ``trick"
that we anticipate will be useful
in other situations.
\end{abstract}


\input{intro}

\input{motivation}

\input{questions}
\input{novelty}
\input{theorems}

\input{theorems2}

\input{conclusions}


\input{ref}
\end{document}

%% file: intro.tex
\section{Introduction}
\label{sec:intro}
A {\em Euclidean Distance Constraint System (EDCS)} 
$(G,\delta)$ is a graph $G = (V,E)$
together with an assignment of  distances $\delta(e)$, or   distance intervals 
$[\delta^l(e),\delta^r(e)]$
to the edges $e \in E$. 
A $d$-dimensional 
{\em realization} is the assignment $p$ of points in $\mathbb{R}^d$
to the vertices in $V$ such that the distance equality (resp.
inequality) constraints 
are satisfied: $\delta(u,v) = \|p(u) - p(v)\|$, respectively
$\delta^l(u,v) \le \|p(u)-p(v)\| \le \delta^r(u,v)$.
Note the EDCS with distance equality constraints, $(G,\delta)$, 
is also called a {\em linkage} and 
was originally referred to as a {\em framework} in combinatorial rigidity
terminology; more recently a framework $(G,p)$ includes a specific
realization $p$, and the distance assignment $\delta$ is read off from $p$.

We seek {\sl efficient representations of the realization space} of 
an EDCS. We define a {\em representation} to be (i) a choice of parameter set, 
specifically a choice of a set $F$ of non-edges of $G$, and 
(ii) a set $\Phi_F^d(G,\delta)$ of possible distance values $\delta^*(f)$ that 
the  non-edges $f\in F\subseteq \overline{E}$ 
can take while ensuring existence of 
at least one $d$-dimensional realization for 
the augmented EDCS: $(G\cup F, \delta(E),\delta^*(F))$.
In the presence of inequalities, the Cayley configuration space is 
denoted  $\Phi_F^d(G, [\delta^l, \delta^r])$ and the 
augmented EDCS is:  
$(G\cup F, [\delta^l(E), \delta^r(E)],\delta^*(F))$.
Here $G \cup F$ refers to a graph $H := (V, E\cup F)$.
In other words, in this manuscript, 
our representations are in {\em Cayley} parameters or non-edge distances: 
the  set $\Phi_F^d(G,\delta)$
(resp. $\Phi_F^d(G, [\delta^l, \delta^r])$)
is the {\sl projection} onto the Cayley parameters in $F$, 
of the Cayley-Menger semi-algebraic set with fixed
$(G,\delta)$ (resp. 
$(G, [\delta^l, \delta^r])$) \cite{bib:blumenthal, bib:menger, bib:cayley}. 
This is also the set of $d$ dimensional 
$|V|\times |V|$ Euclidean
distance matrix completions of the partial distance matrix 
specified by $(G,\delta)$ \cite{bib:alfakih99}.

We refer to the representation 
$\Phi_F^d(G,\delta)$ 
(resp. $\Phi_F^d(G, [\delta^l, \delta^r])$)
as the {\em Cayley configuration space} of the EDCS
$(G,\delta)$ 
(resp. $(G, [\delta^l, \delta^r])$)
{\em on the parameter set $F$} of non-edges of $G$.

\medskip{\bf Note.}
For ease of exposition, from now on EDCS will generally refer to 
distance equality
constraints only. We will indicate with explicit remarks when a theorem
is applicable to EDCS with distance inequality constraints as well. 

The PhD thesis \cite{bib:Gao08} formulates the concept of 
efficient Cayley configuration space description for EDCS
by emphasizing the exact choice of parameters used to represent 
the realization space.
This sets the stage for a {\sl mostly combinatorial, and 
complexity-graded} program of investigation.
An initial sketch of this program was presented in
\cite{bib:GaoSitharam05};  
a comprehensive list of theoretical 
results and applications to date can be 
found in the PhD thesis \cite{bib:Gao08}. 

\subsection{Organization}
In Sections \ref{sec:motivation} and \ref{sec:definition}
we motivate and give a brief background for the overall program of investigation.
The questions of interest and contributions of this manuscript are listed in
Section \ref{sec:contribution}. 
Their novelty and technical significance are outlined in
Section \ref{sec:novelty} together with related work.
Formal results and proofs are presented in Section \ref{sec:results}.
We conclude with suggestions for future work in Section \ref{sec:conclusion}.

%% file: motivation.tex
\section{Motivation}
\label{sec:motivation}
Describing and sampling the realization space of an EDCS 
is a difficult problem  that arises in 
many classical areas of mathematics and theoretical computer science
and has a wide variety of applications in computer aided
design for mechanical engineering, robotics and molecular modeling.
Especially for {\it underconstrained} or {\em independent and not rigid} 
EDCS whose realizations have 
one or more degrees of freedom of motion, progress
on this problem has been very limited.

Existing methods for sampling EDCS realization spaces
often use Cartesian representations, 
factoring out the Euclidean group by 
arbitrarily ``pinning" or ``grounding"
some of the points' coordinate values. 
Even when the methods use ``internal" representation
parameters such as Cayley parameters (non-edges) 
or angles between unconstrained objects, the choice of these 
parameters is adhoc.
    While Euclidean motions may be automatically factored out in the 
    resulting parametrized Cayley configuration space, 
for most such parameter choices, the 
Cayley configuration space is still a topologically complex 
semi-algebraic set, sometimes of reduced measure in high dimensions. 


\begin{figure}
\psfrag{1}{$v_1$}
\psfrag{2}{$v_2$}
\psfrag{3}{$v_3$}
\psfrag{4}{$v_4$}
\psfrag{5}{$v_5$}
\psfrag{x}{$\delta^*(v_2,v_4)$}
\psfrag{y}{$\delta^*(v_1,v_4)$}
\psfrag{d}{$1$}
\psfrag{p00}{$(0,0)$}
\psfrag{p10}{$(1,0)$}
\psfrag{p01}{$(0,1)$}
\psfrag{p12}{$(1,2)$}
\psfrag{p21}{$(2,1)$}
\psfrag{p22}{$(2,2)$}
\centerline{
\includegraphics[width=10cm]{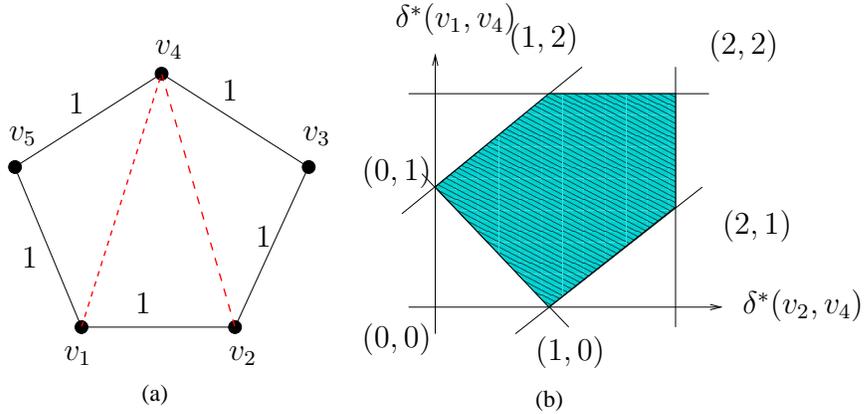}
}
\caption{
When parameters for the EDCS in (a) are chosen to be the two dashed non-edges, 
we get a convex 2D Cayley configuration space shown in (b).
}
\label{F:goodchoice}
\end{figure}

\begin{figure}
\psfrag{1}{$v_1$}
\psfrag{2}{$v_2$}
\psfrag{3}{$v_3$}
\psfrag{4}{$v_4$}
\psfrag{5}{$v_5$}
\psfrag{5'}{$v_{5\p}$}
\psfrag{p1}{$p(v_1)$}
\psfrag{p2}{$p(v_2)$}
\psfrag{p3}{$p(v_3)$}
\psfrag{p4}{$p(v_4)$}
\psfrag{p5}{$p(v_5)$}
\psfrag{p5'}{$p(v_{5\p})$}
\psfrag{a}{$a$}
\psfrag{b}{$b$}
\psfrag{c}{$c$}
\psfrag{d}{$d$}
\psfrag{e}{$\delta^*(v_1,v_3)$}
\centerline{
\includegraphics[width=10cm]{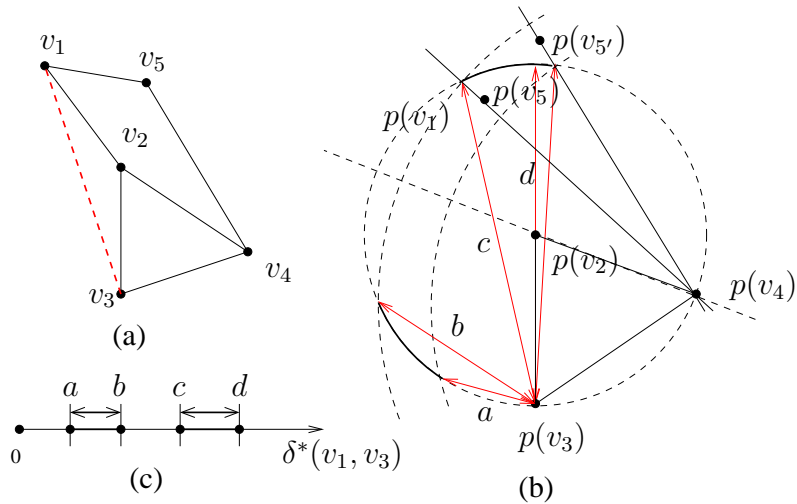}
}
\caption{When parameter for the EDCS in (a) is chosen to be the dashed non-edge, 
we get a disconnected 2D Cayley configuration space shown in (b): the realization  $p(v_1)$ can 
lie in either of the two solid arc segments of the circle labeled $p(v_1)$, yielding
two disconnected intervals for the Cayley configuration space on the non-edge 
$(v_1,v_3)$ as shown in (c).
}
\label{F:badchoice}
\end{figure}


After the representation parameters are chosen, the method of sampling
the Cayley configuration space 
often reduces to ``take a uniform grid sampling and 
throw away sample configurations that do not
satisfy given constraints." 
Since even Cayley configuration spaces of full measure
(representation using lowest possible
number of parameters or dimensions) often have complex boundaries,
potentially with cusps and large holes, this type of sampling 
is likely to  miss extreme and boundary
configurations and is moreover computationally inefficient. 
To deal with this,  numerical, iterative methods are generally used when 
the constraints are equalities, and in the case of inequalities, 
probabilistic ``roadmaps" and
other general collision avoidance methods are used. They
are approximate methods. If the Cayley configuration space is
relatively low dimensional, then  initial
sampling is used to provide an approximate and refinable
representation of the Cayley configuration space, using traditional 
approximation theory
methods such as splines or computational geometry representations,
for example, based on Voronoi diagrams. 
Thereafter this approximate representation
is used to guide more refined sampling. 
All of these 
are approximate methods that do not leverage exact descriptions of the
Cayley configuration space.

Two related problems additionally occur in NMR molecular 
structure determination and wireless sensor network 
localization: completing a partially specified
Euclidean Distance Matrix
in a given dimension; and finding a Euclidean Distance Matrix in a given 
dimension that closely approximates a given 
Metric Matrix (representing pairwise distances in a metric space) 
\cite{bib:yinyuye06,bib:havel88}. 
The latter problem also arises in the study of algorithms for low distortion
embedding of metric spaces into Euclidean spaces of fixed dimension
\cite{bib:indyk06}.
Both of these can be viewed as searching over a Cayley configuration space of an EDCS. 
But the common methods for these problems 
are different from those used for exploring Cayley configuration spaces. 
One reason for this is that 
usually only one realization is usually sought, which optimizes
some appropriately chosen function; the goal is not 
sampling or description of the 
entire Cayley configuration space. Common methods 
for these problems are: 
(i) either use semi-definite programming, since
Euclidean Distance Matrices in a specified dimension 
are directly related to Gram matrices which are 
positive semidefinite matrices of a specified rank; (ii) or 
iteratively enforce the Cayley-Menger determinantal conditions
that characterize Euclidean Distance Matrices in a specified dimension.

\subsection{Exact, efficient Cayley configuration spaces}
\label{sec:definition}
Motivated by these applications, our emphasis
is on {\em exact, efficient} description of the Cayley configuration space of 
{\em underconstrained} or {\em independent and not rigid} EDCS.
(i) An exact algebraic 
description guarantees that boundary and extreme configurations are not missed
during sampling, which is important for many applications.
(ii) An efficient description (i.e, low dimensional, 
full measure, convex, using few polynomial or even linear inequalities,
whose coefficients are obtained efficiently from the given EDCS) 
is important for tractability of the sampling algorithm.

{\sl Efficiency} refers to several factors.
We list four efficiency factors
that are relevant to this manuscript.
The first factor is
the {\em sampling complexity}: 
given the EDCS
$(G,\delta)$, (i) the complexity of computing (ia) the set of Cayley
parameters or non-edges $F$ and (ib) the description of the
Cayley configuration space $\Phi_F^d(G,\delta)$ as a
semi-algebraic set, which includes the algebraic complexity of
the {\sl coefficients} in the polynomial inequalities that describe the
semi-algebraic set,
 and (ii) the descriptive
 algebraic complexity, i.e., number, terms, degree
 etc of the polynomial inequalities that describe the
 semi-algebraic set.
 These together determine the complexity of
 sampling or walking through configurations in
 $\Phi_F^d(G,\delta)$.

Concerning (ia), it is important to note that most choices
of Cayley parameters (non-edges) to represent the realization space
of $(G,\delta)$ give inefficient descriptions of the resulting parametrized
Cayley configuration space
(see for example Figures \ref{F:goodchoice},\ref{F:badchoice}).
Hence we place a strong
emphasis is on a {\sl systematic, combinatorial
choice} of the Cayley parameters  that {\sl guarantee} a
Cayley configuration space with {\sl all} the efficiency requirements
listed here. Further, we are interested in combinatorially characterizing
{\sl for which
graphs $G$ such a choice even exists}.

The second efficiency factor is the {\em realization complexity}.
Note that the price we pay for insisting
on exact and efficient Cayley configuration spaces
is that the map from the traditional Cartesian realization space to the
parametrized Cayley configuration space is many-one.
{\sl I.e, each parametrized or Cayley configuration could correspond to many
(but at least one) Cartesian realizations.} 

However, we circumvent this difficulty by defining and studying 
realization complexity as one of the 
{\sl requirements} on efficient Cayley configuration spaces
i.e., we take into account that the realization step 
typically follows the sampling step, and 
ensure that one or all of the corresponding Cartesian 
realizations can be obtained efficiently from a parametrized
sample configuration.

A third efficiency factor is {\em generic completeness},
i.e, we would like (a) each configuration in our parametrized configuration
space to generically correspond to
at most finitely many Cartesian realizations and (b) we would
like the Cayley configuration space to be of {\em full measure}, and in particular, they 
use {\it at most} as many parameters or dimensions as the 
internal degrees of freedom of $G$.
Specifically (a) means $G\cup F$ is 
rigid and (b) means $G\cup F$ is {\it not overconstrained}, i.e, it is 
{\em independent}.
This generic completeness means that the graph $G\cup F$ is 
{\em wellconstrained}
i.e., {\em minimally rigid}. 

\medskip
{\bf Note.}
In this paper, unless otherwise specified, we always assume that Cayley configuration
spaces have full measure.

A fourth and fifth important efficiency factors are {\em topological 
and geometric complexity}
for example, number of connected components, and convexity. 
Convexity and connectedness are natural properties 
to study since they facilitate convex programming
and other efficient methods for sampling.
Another crucial reason for studying convexity is that 
results (such as those presented here) about convex configuration 
spaces readily generalize 
from Euclidean distance equality constraint  systems
(e.g. frameworks) to Euclidean distance  inequality
constraint systems (e.g. tensegrities and partially specified 
Euclidean Distance Matrices with intervals as entries).

\subsection{Combinatorial Characterization}
Combinatorial characterizations of generic properties of EDCS 
are the cornerstone of combinatorial rigidity theory. In practice 
they crucial for tractable and efficient geometric constraint solving, since  
they are used to analyze and decompose the underlying algebraic system.
So far such characterizations have been used primarily for broad classifications
into well- over- under- constrained, detecting dependent constraints
in overconstrained systems and finding completions for
underconstrained systems. Such combinatorial characterizations have been 
missing in the finer classification of underconstrained
systems according to the efficiency or complexity of their  
Cayley configuration space. 
Our emphasis in this respect is the surprising fact that there is a
clean combinatorial characterization {\sl at all} for the algebraic
complexity of configuration
spaces.
This however is a necessary step in efficiently decomposing and analyzing 
underconstrained systems.

%% file: questions.tex
\section{Questions and Contributions}
\label{sec:contribution}
Next we give 4 natural questions concerning efficient configuration
spaces and the contribution of this manuscript towards answering them.

\subsection{Question 1: Graphs with connected, convex, linear polytope  
 2D Cayley configuration spaces} 

We are interested in characterizing $G$ for which there is a set $F$
of non-edges such that 
for all distance assignments $\delta(E)$ (resp. intervals
$[\delta^l(E), \delta^r(E)]$),
the $d$-dimensional Cayley configuration space 
$\Phi_F^d(G,\delta)$
(resp. $\Phi_F^d(G, [\delta^l, \delta^r])$),
is convex or connected  or is a linear polytope.
Furthemore, given $G$ in this characterized class, we would like to 
characterize the corresponding sets $F$ of non-edges. 
These are exactly the parameter choices that yield efficient
Cayley configuration spaces.

\medskip\noindent
{\bf Note.}
Here the phrase {\em linear} polytope refers to aspect (ib) of the sampling
complexity defined in Section \ref{sec:motivation}:
the coefficients of the linear inequalties that define the bounding hyperplanes 
of the polytope are simple linear combinations of the given $\delta$
determined by $G$ and $F$. Contrast this with a polytope for which 
the coefficients of the bounding linear inequalities are
the solution of an arbitrary polynomial system 
determined by $(G,\delta)$ and $F$.

\begin{itemize}

\item[(1)]
In Theorem \ref{the:mainLooser},
we give an {\sl exact} characterization of the class of graphs $G$ all of whose
corresponding EDCS $(G, \delta)$  admit a 2D {\sl
(generically complete), linear polytope} Cayley configuration space. The theorem also
shows that the characterization remains unchanged if the Cayley configuration space is
merely required to be convex, and further if it is merely required to be
connected.

\item[(2)]
For a graph in the above class, 
in Theorem \ref{the:new-EdgesetProjection} 
 we give an exact combinatorial 
characterization of the choices of Cayley parameters (non-edges) that ensure  
a {\sl (generically complete),
linear polytope} Cayley configuration space.

\item[(3)]
Both above results rely on  key Theorem \ref{the:new-oneEdgeProjection} (in turn based on 
Theorem \ref{the:new-graph2}) that 
characterizes a graph $G$ along with a non-edge $f$ such that 
for all 
distance assignments $\delta(E)$, 
the 2D Cayley configuration space 
$\Phi_f^2(G,\delta)$,
is  a {\sl single interval}.
We additionally show in Observation \ref{obs:3dCounter} 
that this result is tight in that the 
obvious analog of this result fails in 3D.

\item[(4)]
Observation~\ref{obs:interval} shows that 
while the forward direction of 
Theorem \ref{the:mainLooser},
Theorem \ref{the:new-EdgesetProjection}, and 
Theorem \ref{the:new-oneEdgeProjection} for pure distance constraints holds
directly for interval constraints  
$(G, [\delta^l,\delta^r])$, the reverse direction fails. 
However, in Theorem \ref{the:interval},
we give an {\sl exact} characterization of the class of graphs $G$ all of whose
corresponding EDCS $(G, [\delta^l,\delta^r])$ admit a 2D 
(generically complete), linear polytope, convex or connected Cayley configuration space.

\end{itemize}

We note that 
the forward direction of the above theorems (that the graph-theoretic property
always admits a convex, connected, linear polytope Cayley configuration space)
is straightforward. It is the  
reverse direction that is surprising and the proof requires a 
new type of restricted edge-contraction reduction to a graph minor (see 
Section \ref{sec:novelty}).

\subsection{Question 2: Graphs with universally inherent, connected
and convex configuration
spaces}

One can view Contributions (2) and (3) above as characterizing pairs $(G,F)$
such that $G$ always admits a connected or convex Cayley configuration space on $F$.
Sometimes, it is more convenient to instead characterize 
the graphs $H = G\cup F.$ 
In particular, we say that a graph $H$ always admits an {\em inherent} connected
or convex
Cayley configuration space, if  
{\sl there exists} a partition of the edges of $H$ into $E \cup F$ so that
the graph $G=(V,E)$ always admits a connected or convex 
Cayley configuration space on $F$. 
In other words, 
for all distance assignments $\delta(E)$ (resp. intervals
$[\delta^l(E), \delta^r(E)]$) for the 
the graph $G = (V,E)$, the $d$-dimensional Cayley configuration space 
$\Phi_f^d(G,\delta)$
(resp. $\Phi_f^d(G, [\delta^l, \delta^r])$),
is connected or convex. We additionally consider the following strong property.
We say that a graph $H$ always admits {\it universally inherent} connected
or convex Cayley configuration spaces, if  
{\sl for every} partition of the edges of $H$ into $E\cup F$, 
the graph $G=(V,E)$ always admits a connected or convex Cayley configuration space on $F$.
We are interested in combinatorially characterizing 
graphs $H$ that always admit
universally inherent connected or convex Cayley configuration spaces.

\begin{itemize}
\item[(5)]
A graph is $d${\em-realizable} if for every $\delta$ for which 
the EDCS $(G,\delta)$ has a Euclidean realization in any dimension,  
it also has a realization in $\mathbb{R}^d$. 
This useful notion of $d$-realizability was introduced by 
\cite{bib:connelly05,bib:slaugher04}, which also showed a forbidden 
minor characterization of such graphs for $d \le 3$. 
For any dimension $d$, we show in Theorem \ref{the:EDMandRealizability} 
that the class of $d$-realizable
graphs   always admit 
universally inherent connected Cayley configuration spaces, 
that are in fact convex over {\sl squared} Cayley parameters.
We refer to those as  convex {\em squared Cayley configuration spaces}.
This result holds also when the corresponding
EDCS use distance intervals.

\item[(6)]
Theorem \ref{the:2D3D} shows the reverse direction of (5) for 
3D, and
thus shows that 3-realizable graphs 
$H$ are {\sl exactly} the ones  that always 
admit universally inherent connected and convex 3D squared Cayley configuration spaces,
also when the corresponding EDCS use distance intervals.
Thus, by \cite{bib:connelly05,bib:slaugher04}, this class also has 
forbidden minor characterization.
In Observation \ref{obs:3dCounter}, we observe that both 
Theorem \ref{the:EDMandRealizability} 
and 
Theorems \ref{the:2D3D} 
are weak statements 
for 2D -- much stronger statements
follow directly from (2) above.
For example, it follows from (2) that 
if a graph is not 2-realizable, then there is a natural component of 
the graph that  does not even
admit an inherent connected Cayley configuration space description, 
leave alone a universally inherent one.

\end{itemize}

\subsection{Question 3: Efficiently 
recognizing graphs with connected and convex configuration
spaces}

\begin{itemize}
\item[(7)]
Both characterizations in Contributions (1) and (6) 
(Theorems \ref{the:new-EdgesetProjection} and \ref{the:2D3D}) directly point to 
efficient, existing algorithms 
that {\sl recognize} whether a given graph always admits
(universally inherent)  connected, convex, (generically 
complete),  linear polytope 2D and 3D (squared) Cayley configuration spaces.
For example, a recent algorithm for recognizing 3-realizable graphs  
\cite{bib:yinyuye06-2} was given based on a characterization of such graphs
in \cite{bib:connelly05}.
\end{itemize}

\subsection{Question 4: Sampling and Realization complexities
}

The practical use of the contributions (1)-(6) above 
becomes apparent when we answer the following question. 
Given an EDCS $(G,\delta)$ (resp. $(G, [\delta^l,\delta^r])$) 
where $G$ is 
in one of the classes characterized in contributions (1) to (6),
what is the complexity
of computing (i) an appropriate set of Cayley parameters or non-edges $F$ 
(ii) the exact description of
$\Phi_F^d(G,\delta)$ 
(resp. $\Phi_F^d(G, [\delta^l,\delta^r])$) 
as a semi-algebraic set (iii) a cartesian 
realization of a given parametrized configuration 
in $\Phi_F^d(G,\delta)$.
Here (i) and (ii) determine the sampling complexity and (iii) determines
realization complexity.

\begin{itemize}
\item[(8)]
For 2D, Theorem \ref{the:mainLooser} and Theorem~\ref{the:new-EdgesetProjection} shows that the time complexities for 
(i), (ii) and (iii) are linear.
For 3D, our result in Theorem \ref{the:2D3D} 
only pertains to universally inherent 
Cayley configuration spaces, and we only have a weak analog of 
Theorem \ref{the:new-EdgesetProjection}, hence Question 4(i) only marginally applies.
We observe that a result from \cite{bib:yinyuye06-2} gives a weak answer for 
question (i) and  an $O(|E|^4)$ time complexity for (iii).
The complexity for (ii) is an open problem and is discussed in Section
\ref{sec:conclusion}. However, by employing 
Contribution (5) (Theorem~\ref{the:2D3D}), the complexity of obtaining {\sl one} 
configuration is seen to be 
polynomial (in contrast to obtaining a full description of the
Cayley configuration space as a semi-algebraic set).

\end{itemize}

%% file: novelty.tex
\section{Novelty and Related Work}
\label{sec:novelty}

\begin{itemize}

\item 
The study of the cartesian realizations of
plane linkages or EDCS with distance equality constraints 
has a long history \cite{bib:kempe}: over a century
ago, Kempe showed that any plane algebraic curve can be traced by
a point in the realizations of a linkage.
Furthermore, there are extensive studies of the topology of the
configuration  space of,  for example, polygonal linkages 
parametrized by Cayley parameters, see  for example
\cite{bib:hausman}.

However, to the best of our knowledge, ours is the first 
combinatorial or graph-theoretic forbidden minors
characterization of 
geometric and topological complexity such as convexity, and  
connectedness of Cayley configuration spaces in Cayley parameters even 
for the plane, and certainly for 3D. 

Conceptually, the results on global rigidity \cite{bib:hendrickson-uniquerel, bib:jackson05} (resp.
globally linked pairs \cite{bib:jackson06}) are related, since they 
combinatorially characterize when the Cayley configuration space on 
all (resp. specific) Cayley parameters (non-edges)
is a single point. 
However, these characterizations hold only generically \cite{bib:connelly,
bib:Gortler} 
as is customary for many combinatorial properties related to 
rigidity. 
In contrast, we note that our characterizations apply to {\sl all} EDCS'
(frameworks) and not just generic frameworks.
This is a crucial distinction that is needed to reconcile apparent
discrepancies of the two types of results, as we elaborate in 
Section \ref{sec:globally-linked} together with 
a conjecture concerning the modification of our characterization
for generic frameworks.

Moreover, in Section \ref{sec:special-distance} we 
point out that the class of EDCS with convex 
(squared) Cayley configuration spaces is much larger than in 
our characterization,
when the distances
associated with the edges of the EDCS are known to be special.  
Furthermore, in Sections \ref{sec:3dccs} we point out that
the class of EDCS that admit connected 3D Cayley configuration spaces - 
that are not necessarily universally inherent - 
is also much larger than in our characterization.
Both these observations have computational chemistry applications
as pointed out in Sections \ref{sec:helix} and \ref{sec:zeolite}.

Our results additionally yield a combinatorial characterization of
sampling complexity. This incorporates 
the complexity of (i) obtaining the 
chosen set of parameters 
and (ii) the algebraic complexity of obtaining the description of the
Cayley configuration space from the given graph. This in turn includes 
the descriptive complexity of the Cayley configuration space as a 
semi-algebraic set, 
such as the number and degree of the polynomials.  
The characterizations moreover incorporate 
realization complexity, i.e, the complexity of obtaining a realization from a
parametrized configuration.

To the best of our knowledge, the only results of a similar flavor 
are: 
the result of \cite{bib:Owen02}
that shows the equivalence of Tree- or Triangle- decomposability
\cite{bib:Owen91}
and Quadratic or  Radical realizability for planar graphs;
and the result characterizing the sampling complexity of 
1-dof Henneberg-1 graphs \cite{bib:GaoSitharam08b}.

\item
Concerning the use of Cayley parameters or non-edges~
for parametrizing a {\sl generically complete} Cayley configuration space:  
\cite{bib:survey} as well as \cite{bib:JoanArinyo03,bib:ZhangGao06}
study how to obtain ``completions'' of underconstrained graphs $G$, i.e, a set
of non-edges $F$ whose addition makes $G$ minimally rigid or well-constrained.
Both are motivated by realization complexity of underconstrained EDCS: 
i.e, efficiently
obtaining a realization 
{\sl given the parameters values  of a configuration}, i.e,  
once the distance values of the completion edges in $F$ are given.  
In particular \cite{bib:JoanArinyo03} also guarantees that the completion
ensures Tree- or Triangle- decomposability, thereby ensuring low realization
complexity.
However, they do not even attempt to address 
the question of how to find realizable distance {\sl values} for
the completion edges.
Nor do they concern  themselves with 
the geometric, topological or algebraic complexity of 
{\sl the set of distance values
that these completion non-edges can take}, nor the complexity of obtaining 
a description of this Cayley configuration space, 
given the EDCS $(G,\delta)$ and the non-edges
$F$, nor a combinatorial characterization of graphs for which this 
sampling complexity is low.
The latter factors however  are crucial for tractably analyzing and
decomposing underconstrained systems and for 
sampling their Cayley configuration spaces  
{\sl in order to} obtain the corresponding realizations.
The problem has generally been considered too messy, and 
barring effective  heuristics for certain cases, for example in 
\cite{bib:hilderick06},  
there has been no systematic, formal program to study this problem.

\item
Some of the proofs  (e.g. Theorem \ref{the:EDMandRealizability})
employ an unusual interplay of 
(i) classical analytic results related to 
(squared) Euclidean distance matrices, 
such as positive semi-definiteness 
that date back to \cite{bib:schoenberg},
with (ii) recent graph-theroretic characterizations
\cite{bib:connelly05,bib:slaugher04}
related to $d$-realizability. 
This further permits us to directly apply  a recent result
about efficient realization of 3-realizable EDCS \cite{bib:yinyuye06-2}.

\item Some of the proofs, e.g. Theorem \ref{the:new-graph2} 
, use a novel type of restricted edge-contraction reduction
to a graph minor which disallows edge removals a specified pair of vertices to remain distinct and to
remain a non-edge. We anticipate that this new ``trick'' could be useful in other
situations
\cite{bib:trickwiki}.

\end{itemize}

%% file: theorems.tex
\section{Theorems}
\label{sec:results}

\subsection{Basics}
We start with basic definitions and facts.  
Take two graphs $G_{1}$ and $G_{2}$ that both contain a complete graph on 
$k$ vertices, $K_{k}$, as a proper
subgraph. For any matching of the vertices of the two $K_{k}$'s, 
by identifying the matched pairs, we can get a new graph $G_3$.  
We call this procedure a {\em $k$-sum} of $G_1$ and $G_2$
\cite{bib:connelly05}.  A graph is a 
{\em k-tree} if it is a $k$-sum of $K_{k+1}$'s.
Given a graph, we can run the inverse operations of $k$-sum to get
a set of $k$-{\em sum components}. If we can not run the inverse operations of $k$-sum
for a component, we say that component is a {\em minimal ($k$-sum) component}.
Given a non-edge $f$, a {\em minimal $k$-sum component containing $f$}
is a minimal subgraph that is both a $k$-sum component and contains the vertices
of $f$. We emphasize that the phrase does {\em not} refer to a $k$-sum
component that is minimal and happens to contain $f$.

\begin{figure}
\centerline{
\includegraphics[width=6cm]{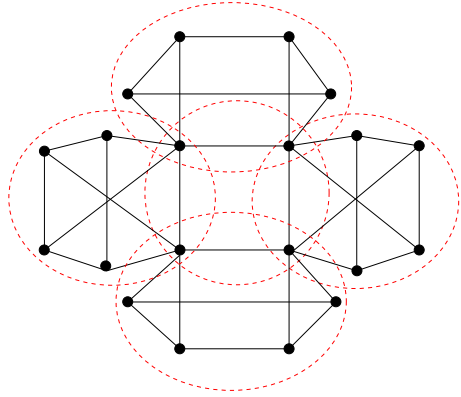}
}
\caption{A 2-sum of five minimal 2-sum components (marked by dashed circles). 
The 2-sum component in
the middle is a partial 2-tree but the entire graph is not a
partial 2-tree. The union of the middle component with any other 
component is also a 2-sum component but not minimal.}
\label{F:2sum}
\end{figure}

 A graph is a
{\em partial k-tree} if it is a subgraph of a $k$-tree. 
Please refer to Figure~\ref{F:2treeAndpartial2tree} for 2-trees and partial 2-trees
and Figure~\ref{F:2sum} for 2-sum and 2-sum component.
It is not hard to see that partial 2-trees are exactly the 2-realizable
graphs. While partial 3-trees are in fact 3-realizable, the class of 
3-realizable graphs include non partial 3-trees as well. 

In \cite{bib:connelly05, bib:slaugher04}
a useful {\em forbidden-minor} characterization of such graphs is given.
A graph $G$ has a graph $K$ as a {\em minor} if there is a vertex 
induced subgraph of $G$ that can be reduced to $K$
via edge removals and edge contractions (coalescing or identifying 
the 2 vertices of an edge).
It is not hard to see that  partial 2-trees are exactly the graphs
that avoid $K_4$ minors. 

Next we give basic 2D combinatorial rigidity definitions 
based on Laman's theorem \cite{bib:Laman70}.
For 3D, no combinatorial definitions exist. 
For the corresponding algebraic definitions, please see for example
\cite{bib:Graver} (combinatorial rigidity terminology) \cite{bib:HoLoSi2001a}
\cite{bib:survey} (geometric constraint solving terminology).

In 2D, a  graph $G=(V,E)$ is
{\em wellconstrained or minimally rigid} if it satisfies the Laman conditions
\cite{bib:Laman70}; i.e.,  $|E|=2|V|-3$ and 
$|E_{s}| \leq 2|V_{s}|-3$ for all subgraphs $G_{s}=(V_{s},E_{s})$ of $G$;
$G$ is {\em underconstrained} or {\em independent and not rigid} if we have $|E| < 2|V|-3$ 
and $|E_{s}| \leq 2|V_{s}|-3$ for all subgraphs $G_{s}$. A graph $G$ 
is {\em overconstrained or dependent} if there is a subgraph $G_{s}=(V_{s},E_{s})$
with $|E_{s}| > 2|V_{s}|-3$. 
$G$ is {\em welloverconstrained or rigid} if there exists a subset of its edges $E'$
such that the graph $G' = (V,E')$  
is wellconstrained or minimally rigid.
A graph is {\em flexible} if it is not rigid.

\subsection{Graphs with connected, convex, linear polytope 2D Cayley configuration space}
We are interested in characterizing graphs $G$ that {\em always admit} a convex,
connected or linear polytope Cayley configuration space; i.e, 
graphs $G$ for which there is a set $F$
of non-edges such that 
for all distance assignments $\delta(E)$ (resp. intervals
$[\delta^l(E), \delta^r(E)]$),
the $d$-dimensional Cayley configuration space 
$\Phi_F^d(G,\delta)$
(resp. $\Phi_F^d(G, [\delta^l, \delta^r])$),
is convex or connected  or is a linear polytope.

Furthemore, given $G$ in this characterized class, we would like to 
characterize the corresponding sets $F$ of non-edges. 
These are exactly the parameter choices that yield well-behaved
Cayley configuration spaces.
\begin{figure}
\centerline{
\includegraphics[width=6cm]{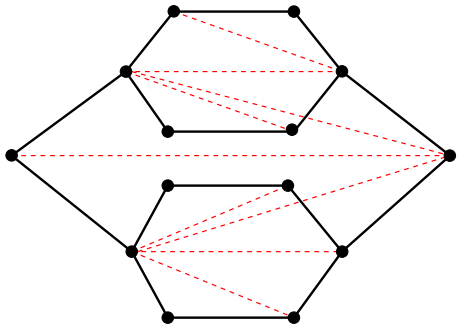}
}
\caption{
The graph of only solid edges is an  underconstrained partial 2-tree
while the graph of both solid and dashed edges is wellconstrained and
is a 2-tree.}
\label{F:2treeAndpartial2tree}
\end{figure}

Figure~\ref{F:2treeAndpartial2tree} gives  an example graph
that admits a connected, convex and linear polytope 2D 
Cayley configuration space on the specified non-edges; and viceversa, 
in Figure~\ref{F:noConnectedComp},
we provide an example in which the graph does {\em not} 
admit such a 
Cayley configuration space on any non-edge. 
The graph characterization 
of Theorem~\ref{the:new-oneEdgeProjection} 
can be easily verified for both examples. 

\subsubsection{Graphs and their ``single-interval'' non-edges}

The next theorem characterizes a graph $G$ along with a non-edge $f$
such that the Cayley configuration space of $G$ on $f$ is always a single interval.

\begin{theorem}
\label{the:new-oneEdgeProjection}
Given a graph $G=(V,E)$ and a non-edge $f$, the
Cayley configuration space  $\Phi_{f}^2(G, \delta)$ is a single interval 
for all $\delta$ if and only if
all the minimal 2-sum components of $G \cup f$ that contain $f$
are partial 2-trees.
\end{theorem}

\begin{figure}
\psfrag{1}{$v_1$}
\psfrag{2}{$v_2$}
\psfrag{3}{$v_3$}
\psfrag{4}{$v_4$}
\psfrag{5}{$v_5$}
\psfrag{6}{$v_6$}
\psfrag{7}{$v_7$}
\psfrag{8}{$v_8$}
\centerline{
\includegraphics[width=6cm]{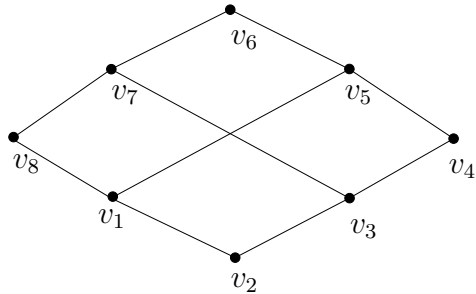}
}
\caption{
No non-edge $f$ exists such that $\Phi_{f}^2(G, \delta)$ is always connected.}
\label{F:noConnectedComp}
\end{figure}

\medskip\noindent
{\sl Structure of Proof:}
To prove one (harder) direction of Theorem~\ref{the:new-oneEdgeProjection}, 
we need the following purely graph-theoretic theorem and the 
following Lemma \ref{lem:basecase}.
The other (easy) direction follows from Lemma \ref{lem:sufficient}
which is in turn 
proven gradually using 
Lemma \ref{lem:2-sum} and 
Lemma \ref{lem:2-sumProjection}.

\begin{figure}
\psfrag{v1}{$v_1$}
\psfrag{v2}{$v_2$}
\psfrag{w1}{$w_1$}
\psfrag{w2}{$w_2$}
\centerline{
\includegraphics[width=6cm]{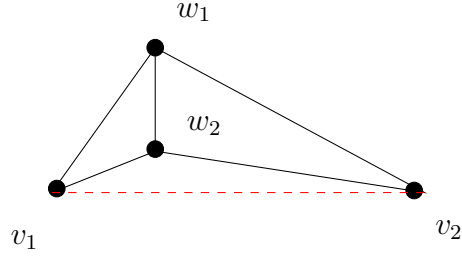}
}
\caption{Base Case 1 of Theorem \ref{the:new-graph2}}
\label{F:base1}
\end{figure}

\begin{figure}
\psfrag{v1}{$v_1$}
\psfrag{v2}{$v_2$}
\psfrag{w1}{$w_1$}
\psfrag{w2}{$w_2$}
\psfrag{u1}{$u_1$}
\psfrag{um}{$u_m$}
\centerline{
\includegraphics[width=6cm]{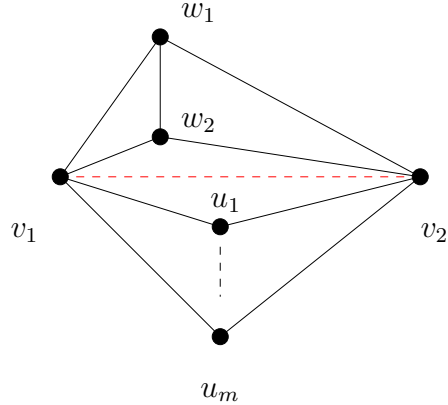}
}
\caption{Base Case 2 of Theorem \ref{the:new-graph2}
: The vertices $u_{i}: i = 1, \cdots, m \mbox{ where m} \geq 1$ 
are the only vertices other than $v_1$ and $v_2$ with degree two and they are adjacent 
to both $v_{1}$ and $v_{2}$;  $f=(v_1,v_2)$ 
is not an edge of the graph}
\label{F:base2}  
\end{figure}

\begin{theorem}
\label{the:new-graph2}
Given graph $G=(V,E)$ and a non-edge $f$,
$G$ can be reduced to the base cases in Figure \ref{F:base1} or Figure \ref{F:base2} 
by a sequence of edge contractions (no edge removals) if and only if
there exists a minimal 2-Sum component of $G \cup f$ containing $f$ that 
is not a partial 2-tree.
\end{theorem}

\begin{remark}
The two base cases Figure \ref{F:base1} and Figure \ref{F:base2} are based on
$K_{4}$. Based on the fact that
partial 2-trees do not  have $K_{4}$ minors and properties of 2-sum, 
we can prove one direction of Theorem~\ref{the:new-graph2}. 
For the other direction the existence of a $K_4$ minor alone is insufficient.
We require a special type of 
pure edge-contraction reduction without edge removals, 
which additionally preserve selected non-edges: i.e, prevent them from becoming edges
or from collapsing to a single vertex.
\end{remark}

\begin{proof} {\bf [Theorem~\ref{the:new-graph2}]}
$(\Rightarrow)$
We first 
prove that $G$ cannot be reduced to Figure \ref{F:base1} or Figure \ref{F:base2}
by edge contractions
if all the minimal 2-Sum components of $G \cup f$ containing $f$ are partial
2-trees. 
Because partial 2-trees do not have $K_{4}$ minors, and  
 since $K_{4}$ exists as a minor in both Figure \ref{F:base1} and Figure \ref{F:base2},
we cannot reduce $G$ to either of the two base cases by edge contractions
(no edge removals).
In fact,  in case there exists  a 2-sum component $G\cup f$
that does not contain $f$, our proof will not change
since edge contractions either preserve 2-sum relationship
or transform a 2-sum to a 1-sum. 


$(\Leftarrow)$
We prove the other direction by induction on the number $n$ of vertices of $G$. 
The statement is true for the 2 base cases.
Assume the statement is true for $|V| \le  n-1$; we prove it for $|V| = n$. 
\begin{figure}
\psfrag{v1}{$v_1$}
\psfrag{v2}{$v_2$}
\psfrag{g1}{$G_1$}
\psfrag{g2}{$G_2$}
\psfrag{gk}{$G_k$}
\psfrag{G}{$G$}
\centerline{ 
\includegraphics[width=6cm]{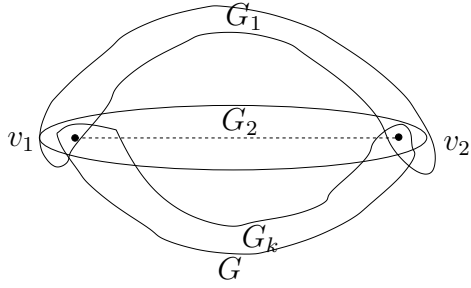}
}
\caption{  
$G$ has a connected 2D Cayley configuration space on $f=(v_1,v_2)$ if and only if
for all $1 \le i \le k$, the graph $G_i$ has a connected 2D Cayley configuration space on $f$.
}
\label{F:connected3d-one-interval}
\end{figure}
First, we remove $v_1$ and $v_2$ to get a set of connected components
$H_{1}, \cdots, H_{k}$  (Figure~\ref{F:connected3d-one-interval}). 
We use $G_i$ to denote the subgraph of $G$ which is induced by vertices of $H_{i}$ 
together with 
$v_{1}$ and $v_{2}$, where $f = (v_1,v_2)$. 
Note that each $G_i \cup f$ is a 2-sum component of $G\cup f$.
Without loss of generality, we assume $G_{1} \cup f$ is one of these  
2-sum components of
$G \cup f$ but not a partial 2-tree.

\smallskip
\noindent
{\bf Case $k > 1$.} If $k > 1$, the number of vertices of $G_{1} \cup f$ is $\le n-1$.
Hence 
by the induction hypothesis, we can reduce $G_{1}$ to the one of the two base cases by 
edge contractions. Now we just contract all the edges of $H_{i}$ when $i > 1$
and the resulting graph falls into the base case in 
Figure~\ref{F:base2}.

\smallskip
\noindent
{\bf Case $k=1$.}
In this case, there is only
one minimal 2-sum component $C$ containing $f$ and it is not a partial 2-tree. 
If $C$
is not $G \cup f$, the number of vertices of $C$ is 
$\le n-1$. By the induction hypothesis we can reduce $C$ to one of the two
base cases. By contracting all the edges of $G$ which are not in $C$,
we can reduce $G$ to one of the base cases. (Note that a 
1-sum is a special case of a 2-sum.)
If, on the other hand,  $C = G\cup f$, then it  is
a minimal 2-sum component containing $f$ and it 
is not a partial 2-tree.
There are 2 subcases based on $l$, 
the maximum number of disjoint paths between $v_{1}$ and $v_{2}$, the
vertices of $f$.
\begin{figure}
\psfrag{v1}{$v_1$}
\psfrag{v2}{$v_2$}
\psfrag{v3}{$v_3$}
\psfrag{g1}{$G_1$}
\psfrag{g2}{$G_2$}
\psfrag{g}{$G$}
\centerline{
\includegraphics[width=6cm]{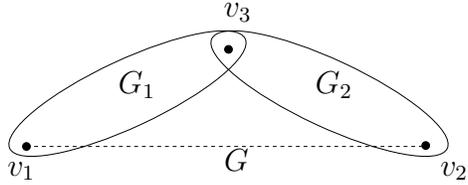}
}
\caption{$v_3$ is an articulation vertex for $v_1$ and $v_2$.}
\label{F:articulationAnd3d-one-interval}
\end{figure}

\smallskip
\noindent
[{\bf Subcase $l\le 1$}] we can find a vertex, say
$v_{3}$, that separates $v_{1}$ and $v_{2}$, that is, we can split the graph into two
subgraphs $G_{1}$ and $G_{2}$ such that $v_{1}\in G_{1}$, $v_{2} \in
G_{2}$; $G_{1}$ and 
$G_{2}$ share only vertex $v_{3}$, and all the edges of
$G$ are either in $G_{1}$ or $G_{2}$ (refer to
Figure~\ref{F:articulationAnd3d-one-interval} for this case). 
Since $G\cup f$ is a minimal 2-sum component containing both $v_1$ and $v_2$, 
both  $(v_1, v_3)$ and  $(v_2, v_3)$ have to be non-edges.
In addition, 
at least one of $G_{1} \cup (v_1, v_3)$ and $G_{2} \cup (v_2, v_3)$ is
not a partial 2-tree, otherwise $G\cup f$ will also be a partial 2-tree.
Without loss of generality, suppose $G_{1} \cup (v_1, v_3)$ is not
a partial 2-tree. By the induction hypothesis, we can reduce $G_{1}$
to one of the two base cases. By contracting all the edges in $G_{2}$ we
can also reduce $G$ to one of the two base cases 
($v_3$ is identified with $v_2$).

\begin{figure}
\psfrag{v1}{$v_1$}
\psfrag{v2}{$v_2$}
\psfrag{t1}{$t_1$}
\psfrag{ts}{$t_s$}
\psfrag{z1}{$z_1$}
\psfrag{zr}{$z_r$}
\centerline{
\includegraphics[width=6cm]{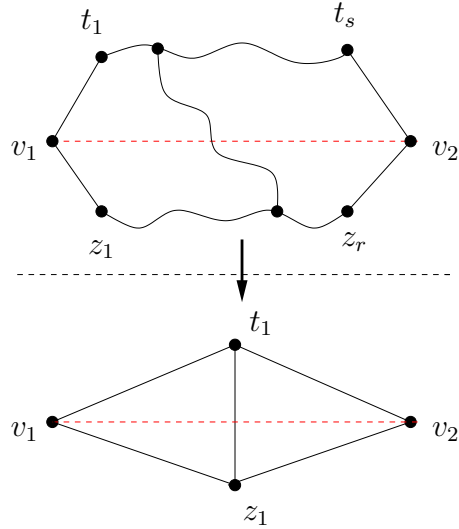}
}
\caption{Case $k=1$, Subcase $l\ge 2$ in proof of Theorem \ref{the:new-graph2}.
There are at least two disjoint paths from $v_{1}$ to $v_{2}$;
$G\cup f$ is
a minimal 2-sum component containing both $v_{1}$ and $v_{2}$; and $G\cup f$ 
is not a partial 2-tree.}
\label{F:proofcase3}
\end{figure}

\smallskip
\noindent
[{\bf Subcase $l\ge 2$}] 
Take two disjoint paths $(v_1,
t_{1},\cdots,t_{s},v_2)$ and $(v_1, z_{1},\cdots,z_{r},v_2)$. We can contract $(
t_{1},\cdots,t_{s})$ to $t_{1}$ and $(z_{1},\cdots,z_{r})$ to $z_{1}$ 
(Figure~\ref{F:proofcase3}). If we
further 
remove $v_1$,$v_2$, $t_{1}$ and $z_{1}$, we get new connected components.
Then we contract all the edges inside these new connected components such that 
each of them becomes a single vertex that we     
denote by $q_{1},\cdots,q_{m}$.  Before we contract paths $(
t_{1},\cdots,t_{s})$ and $(z_{1},\cdots,z_{n})$, if we remove $v_1$ and $v_2$, the 
remaining graph is still connected($k=1$), so at least one of $q_{1},\cdots,q_{m}$
is connected to both $t_{1}$ and $z_{1}$.  


Now we contract edges as follows (refer to Figure \ref{F:proofcase3}).
\begin{enumerate}
\item 
if $q_{i}$ connects to both $t_{1}$ and $z_{1}$, we can identify $q_{i}$ with $t_{1}$ by
edge contraction; 

\item
if $q_{i}$ connects to only $v_1$ and $v_2$ (not to $t_{1}$ or
$z_{1}$), we leave it unchanged;

\item
if $q_{i}$  connects to only one of $v_1$,$v_2$, $t_{1}$ and $z_{1}$, we identify it with the
corresponding vertex in $v_1$,$v_2$, $t_{1}$ and $z_{1}$;  

\item
if $q_{i}$ connects to $v_1$,$v_2$, $t_{1}$, we can identify $q_{i}$ with $t_{1}$;

\item
if $q_{i}$ connects to $v_1$,$v_2$, $z_{1}$, we can identify $q_{i}$ with $z_{1}$.
\end{enumerate}

That covers  all the cases and completes the proof of the induction step.
\end{proof}

\medskip
Theorem~\ref{the:new-graph2} gives us the following independently interesting corollary. 

\begin{corollary}
\label{cor:graph2}
Given graph $G=(V,E)$ and a non-edge $f = (v_1,v_2)$, we
can reduce $G$ to one of the base cases in Figure \ref{F:base1} or 
Figure \ref{F:base2} by a sequence of edge contractions provided the following 
hold. 
\begin{enumerate}
\item $G$ itself is the minimal 2-sum component containing $f$.

\item For any vertex $v_{i}$ other than $v_1$ and $v_2$, either $deg(v_{i})$ is 2 and 
$v_{i}$ is adjacent to both $v_1$ and $v_2$, or $deg(v_{i}) $ is at least three.

\item At least one vertex $v_{i}$ other than $v_1$ and $v_2$ has degree of three or more.
\end{enumerate}
\end{corollary}

\begin{proof}
The proof is by contradiction. Suppose (1), (2), (3) hold but $G$ 
can not be reduced to either of the 
two base cases.
Then by Theorem~\ref{the:new-graph2}, 
all the minimal 2-sum components 
of the graph  $G \cup f$ containing
$v_1$ and $v_2$ are partial 2-trees. Note 
that at least one vertex $v_{i}$ other than $v_1$ and $v_2$ has degree of 
three or more. We  
consider the 2-sum component $C$ containing $v_i$. 
Note also that $C$ has more than 3 vertices, and within $C$, 
since (2) holds,  there can be no vertices
other than $v_1$ and $v_2$ that have degree of two or less. 
By (2), any other vertex of degree two 
or less would be adjacent to both $v_1$ and $v_2$
and would form its own 2-sum components. 

Now note that in any 2-tree with more than 3 vertices, there
exist two non-adjacent vertices which both have degree of two, so in any
partial 2-tree with more than 3 vertices, there should also
exist two non-adjacent vertices which both have degree of at most two. 
This contradicts the
fact that $v_1$
and $v_2$ (which are adjacent) are the only vertices in $C$ that
have degree of two or less. 
\end{proof}

Using Corollary~\ref{cor:graph2}, we see that the 
graph in Figure~\ref{F:noConnectedComp}
can be reduced to one of 
the two base cases no matter which non-edge we choose.

\begin{lemma}
\label{lem:basecase}    
In both Figure \ref{F:base1} and Figure \ref{F:base2}, there exists a
distance assignment $\delta$ s.t. $\Phi^2_f(G, \delta)$ is not connected.
\end{lemma}

\begin{proof} 
Follows from reflection across edge $(w_1,w_2)$. 
Let $\delta(v_1, w_1)$, $\delta(v_1, w_2)$,
$\delta(v_2, w_1)$, $\delta(v_2,w_2)$ and $\delta(w_1,w_2)$ all be 1 
and $\delta(v_1, u_{i})$ and $\delta(v_2, u_{i})$ be 2, we
can easily check that possible values of $\delta(f)$ are 
$0$ or $\sqrt{3}$, so
$\Phi^2_f(G, \delta)$ is not connected. 
\end{proof}

As noted earlier, 
Theorem~\ref{the:new-graph2} and Lemma~\ref{lem:basecase} have proved 
the difficult direction for Theorem~\ref{the:new-oneEdgeProjection}. 
The following lemmas prove the easy direction. 

\begin{lemma}
\label{lem:2-sum}
If $G=(V,E)$ is the 2-sum of $G_{i}=(V_i,E_{i})$,
then for any $\delta$, $(G,\delta)$ has a realization 
if and only if each $(G_{i},\delta)$, 
($\delta$ restricted to the edges in $G_i$)
has a realization.
\end{lemma}

\begin{proof} {\bf [Lemma~\ref{lem:2-sum}]}

Simply hinge all 
the 2-sum components' realizations
along the 2-sum edges to get a realization of
$(G,\delta)$, with any one of two reflection choices across the 2-sum edge. 
The other direction is
immediate.
\end{proof}


\begin{lemma}
\label{lem:2-sumProjection}

Take a graph $G=(V,E)$ with 2-sum components $G_{i}=(v_i,E_{i})$, 
and a non-edge set $F$ that is entirely contained in 
an arbitrary one of the 
$G_{i}=(V_i,E_{i})$, say $G_{1}$. Then for any $\delta$,
either 
$
\Phi_F^2(G,\delta)
=
\Phi_F^2(G_1,\delta)
$ 
if all the 
$\Phi_F^2(G_i,\delta)$'s are non-empty, i.e., the EDCS $(G_i,\delta)$ 
have at least 1 realization; or otherwise,  
$\Phi_F^2(G,\delta)$  is empty.
\end{lemma}

\begin{proof} 

Directly follows from Lemma~\ref{lem:2-sum}.
\end{proof}


\begin{lemma}
\label{lem:sufficient}
\begin{itemize}
\item[(a)] 
If a graph $G=(V,E)$ has a 2-sum component $G' = (V',E')$ that is an
underconstrained partial 2-tree, then there exists a nonempty 
non-edge set $F$ entirely in $G'$ such that for any $\delta$, $\Phi_{F}^2(G,\delta)$
is a linear polytope. Moreover, there is such a set $F$ such that 
$\Phi_F^2(G',\delta)$ is generically
complete for $G'$.

\item[(b)]
If a graph $G=(V,E)$ is an underconstrained partial 2-tree,
then for any nonempty non-edge set $F'$ that preserves
$(V, E \cup F')$ as a partial 2-tree, and 
for all $\delta$,
$\Phi_{F'}^2(G,\delta)$ is 
a linear polytope.
\end{itemize}
\end{lemma}

\begin{proof} 

For proving (a),
we will construct  
a nonempty set of non-edges $F$ in $G'$ and show
that the projection on $F$ is a linear polytope 
for all $\delta$.

Note that $G'$ is an underconstrained partial 2-tree, so we can find a
nonempty subset of non-edges of $G'$
by adding which
we get a 2-tree. We let $F$ be this nonempty set.
Note that such an $F$ is a {\em completion} for $G'$, i.e., 
makes $G'$ minimally rigid.
Hence we know that $\Phi_F^2(G',\delta)$ is of full-measure and generically
complete, proving the last sentence of the theorem.

To get the linear polytope, note that 
a 2-tree can be written as the 2-sum of triangles. 
For example, let 
$\delta(v_i,v_j)$, $\delta(v_j,v_k)$ and $\delta(v_k,v_i)$ denote the
length of the three edges of the triangle $\triangle v_{i}v_{j}v_{k}$, then the   
triangle inequalities are
        $ \delta(v_i,v_j) \leq \delta(v_i,v_k)+\delta(v_j,v_k)$ ,
        $ \delta(v_i,v_k) \leq \delta(v_j,v_k)+\delta(v_i,v_j)$ and
        $ \delta(v_j,v_k) \leq \delta(v_i,v_k)+\delta(v_i,v_j)$ .

Thus, for all $\delta$,  $\Phi_F^2(G',\delta)$ is a linear polytope.
Now since $F$ is entirely in $G'$, 
Lemma \ref{lem:2-sumProjection}
applies  and 
for all $\delta$,
$
\Phi_F^2(G,\delta)
=
\Phi_F^2(G',\delta)
$ 
or
$\Phi_F^2(G',\delta)$  is empty.
Thus, for all $\delta$, $\Phi_F^2(G,\delta)$  
is also a linear polytope. 

For proving (b):
for any  underconstrained 
partial 2-tree $G=(V,E)$,
we can find   
a nonempty non-edge set $F$ that makes
$(V, E \cup F)$ a 2-tree; and we showed that 
for any $\delta$,
$\Phi_F^2(G,\delta)$
 is a generically complete linear polytope 2D Cayley configuration space. 
Take any nonempty subset of $F'$ of such a $F$ - these
are exactly the subsets of non-edges 
whose addition would preserve the partial 2-tree property
of $G$.
Then $\Phi_{F'}^2(G,\delta)$ is the projection of 
$\Phi_F^2(G,\delta)$ on $F'$  and since the latter is a linear polytope,
the former is a linear polytope as well.
\end{proof}

\medskip\noindent
\begin{proof}
{\bf [Theorem \ref{the:new-oneEdgeProjection}]}
The proof of  one (harder) direction follows directly from 
Theorem~\ref{the:new-graph2} 
and the  Lemma \ref{lem:basecase}. 
Specifically, to pick a distance assignment $\delta$ for $G$
that yields a disconnected Cayley configuration space on $f$,
we set all the contracted edges during the procedure of 
Theorem \ref{the:new-graph2} to 0. The uncontracted edges
are now mapped by the reduction to edges of one of the
base cases. Lemma \ref{lem:basecase} tells us how to 
choose those distance values to ensure disconnectedness
of the Cayley configuration space on $f$.
The other (easy) direction
is immediate from Lemma \ref{lem:sufficient}. 
\end{proof}

Next we show that Theorem \ref{the:new-oneEdgeProjection}
is tight in that neither of the two straightforward extensions to 3D hold.

\begin{figure}
\psfrag{1}{$v_1$}
\psfrag{2}{$v_2$}
\psfrag{3}{$v_3$}
\psfrag{4}{$v_4$}
\psfrag{5}{$v_5$}
\psfrag{6}{$v_6$}
\psfrag{f}{$f$}
\centerline{
\includegraphics[width=6cm]{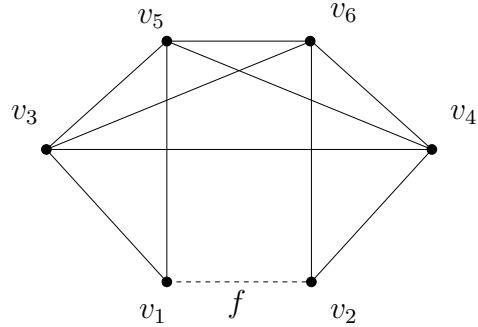}
}
\caption{For Observation 
\ref{obs:3dCounter},
$G$'s  3D Cayley configuration space on non-edge $f$ is one interval, although $G
\cup \{f\}$ has a $K_{5}$ minor.}
\label{F:k5CounterExample}
\end{figure}

\begin{figure} 
\psfrag{1}{$v_1$}
\psfrag{2}{$v_2$}
\psfrag{3}{$v_3$}
\psfrag{4}{$v_4$}
\psfrag{5}{$v_5$}
\psfrag{6}{$v_6$}
\psfrag{7}{$v_7$}
\psfrag{f}{$f$}
\centerline{
\includegraphics[width=6cm]
{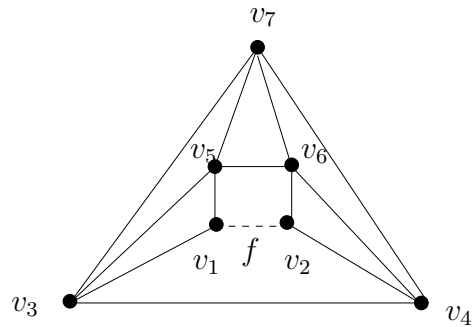}
}
\caption{For Observation 
\ref {obs:3dCounter}.
Graph $G$ has connected 3D Cayley configuration space on non-edge $f$;
$G$ does not
have $K_{5}$ minor or $K_{2,2,2}$ minor; $G \cup
f$ has a $K_{2,2,2}$ minor but does not have a $K_{5}$ minor;
in particular,
without contracting edge $f$ in $G \cup f$ we cannot get a $K_{2,2,2}$
minor.}
\label{F:k222CounterExample} \end{figure}
 
\begin{observation}
\label{obs:3dCounter}
There exists a
partial 3-tree(resp. 3-realizable graph) $G=(V,E)$ and non-edge $f$ such that $G \cup f$ is
not a partial 3-tree(resp. 3-realizable graph) and yet $\Phi_{f}^3(G,\delta)$
is always connected.
\end{observation}

\begin{proof}
Refer to and  Figure~\ref{F:k5CounterExample} and 
Figure~\ref{F:k222CounterExample}.
\end{proof}

\subsubsection{Graphs with generically complete, linear polytope Cayley configuration spaces}

In Theorem \ref{the:mainLooser},
we give an  exact characterization of the class of graphs $G$,  all of whose
corresponding EDCS $(G, \delta)$  admit a 2D {\sl
(generically complete), linear polytope} Cayley configuration space. The theorem also
shows that the characterization remains unchanged if the Cayley configuration space is
merely required to be convex, and further if it is merely required to be
connected.

\begin{theorem}
\label{the:mainLooser}
(a)For a graph $G=(V,E)$,  the following four statements are
equivalent:

\begin{enumerate}
\item There exists a non-empty set of non-edges $F$ such that  for all $\delta$
 $\Phi_{F}^2(G,\delta_E)$ is connected;

\item There exists a non-empty set of non-edges $F$ 
such that for all $\delta$, $\Phi_{F}^2(G,\delta_E)$ is  convex; 

\item There exists a non-empty set of non-edges $F$ such that  for all $\delta$
$\Phi_{F}^2(G,\delta_E)$ is a linear polytope.

\item $G$ has a 2-sum component that is an underconstrained partial
2-tree.
\end{enumerate}
(b)An underconstrained graph $G$ always admits a generically complete linear polytope, connected or
convex Cayley configuration space if
and only if every underconstrained 2-sum component of $G$ is a partial
2-tree. 
\end{theorem}

\begin{proof}
For (a), we will prove the cycle $(4) \Rightarrow (3) \Rightarrow (2) \Rightarrow (1) 
\Rightarrow (4)$.  We proved $(4) \Rightarrow (3)$ in Lemma~\ref{lem:sufficient}. A
linear polytope is  convex, so $(3) \Rightarrow (2)$ follows.  Convexity
implies connectedness, so $(2) \Rightarrow (1)$ follows.  
Theorem~\ref{the:new-oneEdgeProjection}  and the proof of Lemma \ref{lem:sufficient} proves
$\overline{(4)} \Rightarrow \overline{(1)}$, therefore, we proved $(1) \Rightarrow
(4)$ as well. 

For one direction of (b): 
if every underconstrained 2-sum component of $G$ is an underconstrained partial 2-tree, 
then by  Lemma \ref{lem:sufficient}, $G$ always admits a generically complete linear polytope, connected or
convex Cayley configuration space. 
The reverse direction of (b) follows from  (a) (1,2,3 $\Rightarrow$ 4): 
if $G$ always admits a generically complete linear polytope, connected or
convex Cayley configuration space, $G$ has at least one 2-sum component 
which is an underconstrained partial
2-tree. 

\end{proof}

%% file: theorems2.tex
\subsubsection{Full characterization of Cayley parameters that yield a linear polytope 2D Cayley configuration space}
The following theorem is a refined version of  
Theorem \ref{the:mainLooser}.

\begin{theorem}
\label{the:new-EdgesetProjection}
Given a graph $G=(V,E)$ and non-empty set of non-edges $F$, 
the 2D Cayley configuration space $\Phi_F^2(G,\delta)$ is a linear polytope, connected or
convex for all $\delta$ if and only if
all the minimal 2-Sum components of $G \cup F$ containing any subset of $F$ are partial
2-Trees.
Furthermore, the Cayley configuration space is generically complete if and only if
all the underconstrained minimal 2-sum components of $G$ are partial 2-trees and 
all the minimal 2-sum components of $G\cup F$ containing $F$ are 2-trees.
\end{theorem}

\begin{proof}
Directly from Theorem \ref{the:new-oneEdgeProjection} and the proof of Lemma~\ref{lem:sufficient}.
\end{proof}

\subsubsection{Characterization of EDCS with distance intervals}

We have characterized graphs $G$ and sets of non-edges $F$
where the Cayley configuration space
$\Phi_F^d(G,\delta)$ is connected, convex and a linear polytope
for all $\delta$. We can extend
the results for distance interval constraints. 
The next observation and theorem show that while one direction 
of the above results extends to interval constraints, 
the other direction fails in the current form. 
This motivates a modified characterization theorem.

\begin{observation}
\label{obs:interval}
There exists a graph $G=(V,E)$ and non-edge $f$ such that
all the minimal 2-Sum components of $G \cup f$ containing $f$ are partial
2-trees and yet we can find an interval distance constraint $[\delta^l, \delta^r]$
such that $\Phi_f^2(G,[\delta^l, \delta^r])$  is not linear polytope
convex, or
connected.
\end{observation}

\begin{figure}
\psfrag{1}{$v_1$}
\psfrag{2}{$v_2$}
\psfrag{3}{$v_3$}
\psfrag{4}{$v_4$}
\psfrag{5}{$v_5$}
\psfrag{6}{$v_6$}
\psfrag{7}{$v_7$}
\psfrag{8}{$v_8$}
\psfrag{9}{$v_9$} 
\psfrag{10}{$v_{10}$}
\psfrag{11}{$v_{11}$}
\centerline{
\includegraphics[width=10cm]{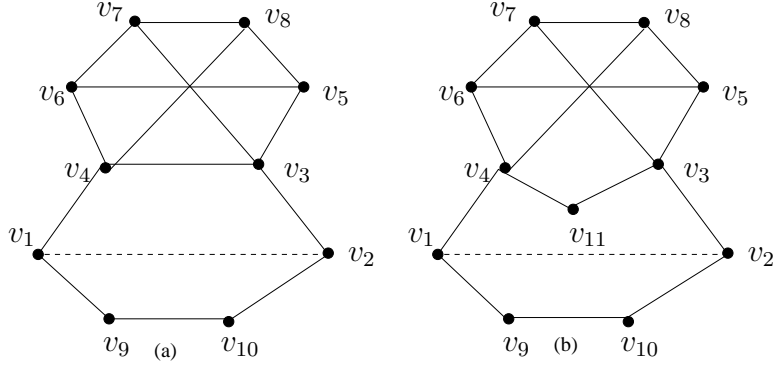}
}
\caption{
For Observation \ref{obs:interval}.
All the minimal 2-sum components of $G \cup f$ containing $f$ are partial
2-trees but $\Phi_f^2(G,[\delta^l(E), \delta^r(E)])$  is not always connected.
}
\label{F:interval} 
\end{figure}

\begin{proof}
Refer to Figure~\ref{F:interval}. Denote the graph  shown in
Figure~\ref{F:interval} (a) by $G$
and the graph  shown in
Figure~\ref{F:interval} (a) by $G'$. 
Denote non-edge $(v_1, v_2)$
by $f$ in both  Figure~\ref{F:interval} (a) and (b). Clearly, $G \cup f$ has 3 minimal 2-sum components:
the subgraph induced by $(v_1, v_2,v_9, v_{10})$,
the subgraph induced by $(v_1, v_2,v_3, v_{4})$
and
the subgraph induced by $(v_1, v_2,v_3, v_{4}, v_5, v_6, v_7, v_8)$.
The first two minimal 2-sum components both contain $f$ 
and both are partial 2-trees.
The
third one (the subgraph induced by $(v_3, v_{4}, v_5, v_6, v_7, v_8)$) 
is not a partial 2-tree but does not contain $f$. 
By Theorem~\ref{the:new-oneEdgeProjection}, 
for all distance $\delta$, 
the Cayley configuration space $\Phi_f^2(G,\delta)$ is connected.
Similarly, 
$G' \cup f$ has 2 minimal 2-sum components:
the subgraph induced by $(v_1, v_2,v_9, v_{10})$,
and
the subgraph induced by $(v_1, v_2,v_3, v_{4}, v_5, v_6, v_7, v_8)$.
Both two minimal 2-sum components contain $f$.
The first one is not a partial 2-tree and the
second one (the subgraph induced by $(v_1, v_2,v_3, v_{4}, v_5, v_6, v_7, v_8)$) 
is not a partial 2-tree. 
By Theorem~\ref{the:new-oneEdgeProjection}, 
we can find distance $\delta'$ such that  
the Cayley configuration space $\Phi_f^2(G,\delta')$ is not connected.
If we allow an interval distance constraint on edge $(v_3, v_4)$ in $G$, 
we would have an equivalent distance constraint sytem as $G'$. 
\end{proof}

Now we give a characterization theorem for distance interval constraints.

\begin{theorem}
\label{the:interval}
Given a connected graph $G=(V,E)$ and a nonempty set of non-edges $F$, 
the 2D Cayley configuration space $\Phi_F^2(G,[\delta^l, \delta^r])$ is linear polytope, convex or
connected for all $[\delta^l, \delta^r]$ if and only if
every minimal 2-Sum component of a specific subdivision $G_F'$ of $G \cup F$ 
that contains any subset of $F$ is a partial
2-tree. We construct the subdivision $G_F'$ of $G \cup F$ by replacing each edge 
$(u_1,u_2)$ by a path of length 2 by introducing a new vertex $u.$ 
\end{theorem}

\begin{proof}
For a given interval constraint $[\delta^l(u_1, u_2), \delta^r(u_1,u_2)]$, 
 we can choose $\delta(u_1,u)= \frac{|\delta^r(u_1,u_2) -\delta^l(u_1,v_2)|}{2}$  
 and $\delta(u,u_2)= \frac{\delta^r(u_1,u_2) + \delta^l(u_1,u_2)}{2}$
 such that the distance constraints on the subdivision, namely 
$\delta(u_1,u)$ and $\delta(u,u_2)$ together impose the original 
distance
interval constraint on the edge $(u_1,u_2)$.
Then by Theorem~\ref{the:new-EdgesetProjection}, $\Phi_F^2(G\p, \delta)$
is a linear polytope, connected or
convex for all $\delta$ if and only if
all the minimal 2-Sum components of $G'_F$ containing 
a subset of $F$ are partial
2-Trees. 
\end{proof}

\subsection{Graphs with universally inherent, convex, squared Cayley configuration spaces}

\subsubsection{$d$-realizability implies convex, squared Cayley configuration spaces}

We will prove in Theorem 
\ref{the:EDMandRealizability}
that $d$-realizable graphs admit universally inherent, connected
(resp. convex) $d$-dimensional (resp. squared) Cayley configuration spaces. Before that, we first
show in Lemma \ref{lem:ConvexAndConnected}
that convexity of squared Cayley configuration spaces implies connectedness
of the Cayley configuration space.

\medskip\noindent
{\bf Note.} All the results in the section apply to both distance constraints $\delta$ as
well as distance interval constraints $[\delta^l,\delta^r]$. In order to 
avoid  writing: ``for all $\delta$(resp. $[\delta^l,\delta^r]$)
the (resp. squared) Cayley configuration space $\Phi_F^d(G, \delta)$ (resp. $\Phi_F^2([\delta^l,\delta^r])$)  
is connected (resp. convex)" we instead just say ``$G$ always admits a
connected (resp. convex) (resp. squared)  Cayley configuration space".

\begin{lemma}
\label{lem:ConvexAndConnected}
If a graph always
admits universally inherent, convex, $d$-dimensional squared Cayley configuration spaces,
 then it also always admits universally inherent, connected, 
$d$-dimensional configuration
spaces. 
\end{lemma}

\begin{proof}
For a non-edge set  $F = \{f_1,f_2,\ldots\}$ of $G$,
denote 
by $(\Phi_F^d)^2(G,\delta)$
the Cayley configuration space: $\{((\delta^*)^2(f_1),(\delta^*)^2(f_2)\ldots)
: (G\cup F,\delta,\delta^*(F))$ has a realization in $\mathbb{R}^d\}$.
The map $(.)^2: \Phi_F^d(G,\delta) \rightarrow (\Phi_F^d)^2(G,\delta)$
is continuous and the inverse map is well-defined and continuous 
over the positive reals.
Now the convexity of 
$(\Phi_F^d)^2(G,\delta)$ implies its connectedness, that by the 
above-mentioned continuity 
implies the connectedness of 
$\Phi_F^d(G,\delta)$. 

\end{proof}

Now we are ready to give the theorem and the proof.

\begin{theorem}
\label{the:EDMandRealizability}
If a graph  is $d$-realizable, it
admits universally inherent, connected (resp. convex), 
$d$-dimensional (resp. squared) Cayley configuration spaces. 
\end{theorem}

\begin{proof}
By Lemma~\ref{lem:ConvexAndConnected}, 
we only need to prove a
$d$-realizable graph admits universally inherent, convex,
squared Cayley configuration spaces.

A $n\times n$ matrix $M$ 
is a {\em Euclidean square distance matrix (EDM)} 
if $\exists p_{1},\ldots, p_{n} \in \mathbb{R}^d$
for some $d$ such that $||p_i -p_j||^2 = M(i,j)$.
A classical result \cite{bib:schoenberg} that follows from
 positive semidefiniteness of Gram matrices is that 
the set of all
EDM's is a convex cone (note that $d$, and hence the rank of 
these matrices is not fixed).
The projection of this cone on 
any set $E\cup F$  of pairs $(i,j)$ is also convex.
By the definition of $d$-realizability of a graph $H = (V,E\cup F)$,
with $|V| = n$, 
this projection is exactly the 
set of all squared distance assignments $(\delta^*)^2$ to the pairs in 
$E\cup F$ for which 
$(H,\delta^*)$ has a realization  in $\mathbb{R}^d$.
We denote this 
$(\Phi_{E\cup F}^d)^2((V,\phi),\phi)$.
Since convexity is preserved by both sections and projections,
the section of this projection 
$(\Phi_{E\cup F}^d)^2((V,\phi),\phi)$ 
- obtained by fixing $\delta^*$ to be 
$\delta$  over $E$ - is also convex, for all $\delta$.
This section is exactly the Cayley configuration space  
$(\Phi_F^d)^2(G, \delta)$ of the graph  $G := (V,E)$
over its nonedge set $F$.  Hence this is a convex squared Cayley configuration space.
Note that this holds for any partition $E\cup F$ of
the edge set of the $d$-realizable graph $H$. Hence $H$ always
admits universally inherent, convex, squared Cayley configuration spaces.
\end{proof}

Theorem~\ref{the:EDMandRealizability} gives one direction 
for all dimensions. 
We conjecture in 
Section~\ref{subsec:conjecture1} 
the reverse direction is also true.
In the next section, we will
prove that the
reverse direction is true for $d\le 3$.

\subsubsection{When does universally inherent, connected configuration 
space imply\\ 
$d$-realizability?}


\begin{theorem}
\label{the:2D3D} 
Let $d \le 3$. Then
the following are equivalent for a graph $H$.
\begin{enumerate}
\item
$H$ is $d$-realizable.
\item
$H$ always admits universally inherent, connected, $d$-dimensional
Cayley configuration spaces.
\item
$H$ always admits universally inherent, convex,  $d$-dimensional
squared
Cayley configuration spaces.
\end{enumerate}
\end{theorem}

\begin{remark}
The above theorem is a weak statement for $d\le 2$.
For example, since 
the class of 2-realizable graphs is exactly the partial 2-trees,
Theorem \ref{the:new-EdgesetProjection} shows that if a graph $H$ is not 2-realizable,
then it has a minimal 
2-sum component that is not 2-realizable (not a partial 2-tree). 
And this component
does not admit {\sl any} 
inherent Cayley configuration space, let alone universally inherent
ones. In other words, in this non 2-realizable minimal 2-sum component $H_C$,
on a vertex set $V_C$
{\em for every} partition of edges into $E_C\cup F_C$, 
the Cayley configuration space 
$\Phi_{F_C}^2(G_C, \delta)$
of graph $G_C  = (V_C,E_C)$
is disconnected. For $d = 3$, on the other hand, no such strong
statement holds as shown in the counterexample of Observation \ref{obs:3dCounter}.
To show the above theorem, we merely show that if a graph $H$ is not
3-realizable, then {\em there exists} a partition of the edges of $H$ 
into $E\cup F$, such that the configuration  space 
$\Phi_{F}^3(G, \delta)$ 
of the graph $G = (V,E)$
is disconnected. 
\end{remark}

\medskip
\noindent
{\sl Structure of Proof.}
The proof of the above theorem requires several lemmas.
The idea of the proof is as follows.
Lemma \ref{lem:ConvexAndConnected} proves (3)$\Rightarrow$(2), and 
Theorem \ref{the:EDMandRealizability} 
proves (1)$\Rightarrow$(3) for all dimensions $d$.
Based on the above remark, we restrict ourselves to $d=3$, and 
just prove (2)$\Rightarrow$(1).

For any non-3-realizable graph $H = (V,E')$,  we 
find a partition of $E'$ as $E\cup F$ where
$G$ is the graph $(V, E)$ 
and find a distance assignment $\delta$ such that 
the Cayley configuration space
$\Phi_F^3(G,\delta)$ is disconnected. 
Here $G := (V,E)$ and $f$ is a single non-edge of $G$,
so this is a 1-parameter Cayley configuration space and we show
that it has 2 isolated points.
To do this, we  start from the following theorem.

\begin{theorem}
[{\bf Connelly and Sloughter \cite{bib:connelly05, bib:slaugher04}}]
\label{the:Connelly}
A graph is 3-realizable if and only if it does not 
have $K_{5}$ or $K_{2,2,2}$
as minors.
\end{theorem}

Given the existence of one of these 2 minors, 
as in Theorem \ref{the:new-graph2},
we show how to pick from $H$ the graph $G$ and its non-edges $F$ such that  
by a restricted reduction that uses 
only edge contractions (no edge removals) and 
preserving the non-edges $F$,
we can reduce  the graph $G$ to a $K_{5}$ or $K_{2,2,2}$ that is 
missing exactly one edge, namely $f$, on to which all the
non-edges in $F$ have been mapped by the reduction. 
Finally, we obtain using Lemma~\ref{lem:k222NotOneInterval} the distance assignment $\delta$ for $G$
by setting the distances for the contracted edges to 0; 
and, similar to Lemma \ref{lem:basecase} 
we pick distance assignments for the un-contracted edges 
in such a way that the two base cases do not have 
connected 3D Cayley configuration spaces on $f$. 

\medskip\noindent
We will use the following simple fact.

\begin{fact}
\label{fact:CliqueMinor}
If a graph $G=(V,E)$ has a complete graph $K_{m}$, as a minor, then $G$ can be
reduced to $K_{m}$ by edge contraction alone (and removal of isolated   
vertices if necessary), without edge removals.
\end{fact}


We also need the following independently interesting theorem
which strengthens 
the forbidden-minor theorem of 
Theorem \ref{the:Connelly}.

\begin{theorem}
\label{the:3dRealizableContraction}

If a graph is not 3-realizable, it can be reduced to $K_{5}$ or
$K_{2,2,2}$  by edge contractions alone (no edge removals).

\end{theorem}

\begin{proof}

If a graph is not 3-realizable, it has a $K_{5}$ or $K_{2,2,2}$ minor 
by Theorem \ref{the:Connelly}. If a graph has a $K_5$ minor, 
use Fact~\ref{fact:CliqueMinor}, it can be reduced to $K_5$ by 
edge contractions alone, so we only need to prove the case that
$G$ has a $K_{2,2,2}$ as minor.
If $G$ has a $K_{2,2,2}$ as minor, by definition of minor we can get a
$K_{2,2,2}$ by first contracting some edges, then removing some edges, 
and finally removing some isolated vertices. Follow this reduction path
but stop after edge contractions, and denote the new
graph by $G'$. Denote the $K_{2,2,2}$ subgraph of $G'$ by
$M$.

Now we will show we can either get $M$ or a $K_{5}$ by edge-contractions
and removing some isolated vertices if necessary.
The strategy is straight forward: 
successively contract one edge at a time, whose two vertices are
not both in $M$ until we cannot continue. After removing any
possible isolated vertices, the remaining graph has exactly 6 vertices(the
same as $M$) and we denoted this graph by $M'$. We know that
$K_{2,2,2}$ is a subgraph of $M'$. Now we use a  simple observation 
that by adding 
one or more edges to $K_{2,2,2}$, we can get a $K_{5}$ by edge
contractions alone, thus $G$ can be reduced to $K_{5}$ or $K_{2,2,2}$ by edge
contractions alone.

\end{proof}

For proving the next lemma, we give  
an appropriate distance assignment $\delta$ 
to $K_{5}$ and $K_{2,2,2}$ to show
they do not admit universally inherent, connected Cayley configuration spaces.

\begin{figure}
\psfrag{1}{$v_1$}
\psfrag{2}{$v_2$}
\psfrag{3}{$v_3$}
\psfrag{4}{$v_4$}
\psfrag{5}{$v_5$}
\psfrag{6}{$v_6$}
\psfrag{7}{$v\p_5$}
\centerline{
\includegraphics[width=10cm]{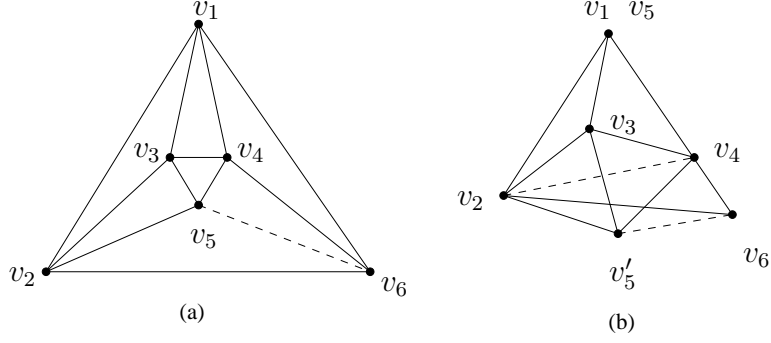}
}
\caption{(a) $K_{2,2,2}$ with one edge $f$ removed; (b) see proof of Lemma~\ref{lem:k222NotOneInterval}: 
a distance assignment $\delta$ to $K_{222} \setminus f$ such
that the 3D Cayley configuration space on $f$ is not connected.}
\label{F:k222DisAssig}
\end{figure}

\begin{lemma}
\label{lem:k222NotOneInterval}
Neither $K_{2,2,2}$ nor $K_5$
always admit universally inherent, connected, 3D Cayley configuration spaces.
\end{lemma}

\begin{proof}
To prove the Lemma for $K_5$ (resp. $K_{2,2,2}$), we take one edge as $f$ 
and prove that we can find a distance assignment $\delta$
for graph $K_5\setminus f$ (resp. $K_{2,2,2}\setminus f$) such that 
$\Phi_{f}^3(K_5\setminus f,\delta)$  (resp. $\Phi_{f}^3(K_{2,2,2}\setminus f,\delta)$) has more than one interval.

For $K_{5}$ case, we take $f$ to be $(v_1, v_2)$
and assign the same distances to all the edges in $K_5\setminus f$.
If we fix tetrahedron $(v_1, v_3, v_4, v_5)$, then $v_4$ can either
be coincident with $v_1$ or is the reflection of $v_1$ about plane
$(v_3, v_4, v_5)$. Since $v_1$  is not in the plane $(v_3, v_4, v_5)$,
so in the latter case, $\delta^*(f)$ is not zero. In the former case,
 $\delta^*(f)$  is zero. These two values are all the possible values
of  $\delta^*(f)$, so we have proved that $\Phi_{f}^3(K_5\setminus f,\delta)$ 
is not connected.


For $K_{2,2,2}$ case (Figure~\ref{F:k222DisAssig}), we choose
edge $(v_5, v_6)$ as $f$. We choose a distance assignment $\delta$ so that the following
conditions are satisfied: $\delta(v_1,v_2)=\delta(v_2,v_3)
=\delta(v_1,v_3)=\delta(v_1,v_4)=\delta(v_3,v_4)=\delta(v_2,v_5)=\delta(v_3,v_6)=\delta(v_4,v_5)$,
$\delta(v_1,v_4)+\delta(v_4,v_6)=\delta(v_1,v_6)$, $\delta(v_4,v_6)>0$, and $\delta(v_2,v_6)$ will 
let $\angle(v_2,v_1,v_6)=\frac{\pi}{3}$.

Because $\delta(v_1,v_4)+\delta(v_4,v_6)=\delta(v_1,v_6)$, $v_{1}$, $v_4$ and $v_6$ are collinear and the four
vertices $v_{1}$, $v_{2}$, $v_4$ and $v_6$ are coplanar. Because 
$\angle(v_2,v_1,v_6)=\frac{\pi}{3}$ and $\delta(v_1,v_2)=\delta(v_1,v_4)$, $\delta(v_2,v_4)$ will be equal to
$\delta(v_1,v_2)$ and
$\delta(v_1, v_4)$. Thus, the location of $v_3$ will be uniquely determined by 
the normal
tetrahedron $(v_1, v_2, v_3, v_4)$ if we assume that, without loss of generality, $v_3$ is
above plane $(v_1, v_2, v_4, v_6)$. Now $v_5$ has two possible locations, either $v_1$ or the
reflection of $v_1$ about plane $(v_2, v_3, v_4)$. We denote the former location as
$v_{5}$ while the latter as $v\p_5$. For the case when $v_5$ is coincident with
$v_1$, $\delta(v_5,v_6)$ will be $\delta(v_5, v_6)=\delta(v_1, v_6)=\delta(v_1, v_4)+ \delta(v_4, v_6)$. For the other case,
by the triangle inequality in $\triangle (v_4, v\p_5, v_6)$, we have
$\delta(v\p_5, v_6) \leq \delta(v\p_5, v_4)+\delta(v_4, v_6) = \delta(v_1, v_4)+ \delta(v_4, v_6)= \delta(v_1, v_6)=\delta(v_5, v_6)$. Now
$(v_1, v_2, v_3, v_4)$ and $(v_2, v_3, v_4, v\p_5)$ are normal tetrahedrons and $v\p_5$ is different
from $v_{1}$ in this case, so $v\p_5$  is not in the plane $(v_1, v_2, v_4)$. Therefore,
$v\p_5$  can not be collinear with $v_4$ and $v_6$, so $\delta(v\p_5, v_6)$ 
$\delta(v_5, v_6)$. Thus $\delta(f)$ has two disconnected values and we
have 
proved that projection $\Phi^3_{f}(G,\delta)$ is not connected.
\end{proof}

Now we are ready to prove Theorem~\ref{the:2D3D}.

\noindent
\begin{proof} {\bf [Theorem~\ref{the:2D3D}]}
By Theorem~\ref{the:EDMandRealizability}, a 3-realizable graph 
$H$
admits universally inherent connected, convex,
3D squared Cayley configuration spaces, so we only need to prove the reverse
direction.
If $H$ is not 3-realizable, by
Theorem~\ref{the:3dRealizableContraction}, we
can get a $K_{5}$ or $K_{2,2,2}$ by edge contractions alone. 

As mentioned before, find a partition of the edge set of $H$ into $F\cup E$
that defines a subgraph $G = (V,E)$ and a 
non-edge set $F$ for $G$. Then find a distance assignment $\delta$ 
for $E$ such that
$\Phi_F^3(G,\delta)$ is disconnected.

Since $H$ can be reduced to $K_{5}$ or $K_{2,2,2}$
by edge contractions alone, we 
 pick an edge from the corresponding minor and denote it $f$. 
We choose $F$ to be all the edges that were identified with $f$
by the reduction.
For the distance assignment $\delta$ to the edges of $G$, 
we will use a similar method as in the proof of Theorem \ref{the:new-oneEdgeProjection}.
Lemma~\ref{lem:k222NotOneInterval} gives a distance assignment
for $K_{2,2,2}$ and $K_5$ 
that ensures that the Cayley configuration space on the edge $f$ is 
disconnected.
Set the distances of each uncontracted edge $e$ during the reduction of $G$ to
the distance assignment of the edge in $K_5$ or 
$K_{2,2,2}$ that $e$ was identified with.
Set the distances of all the contracted edges of $G$ to be 0. 
This ensures that 
$\Phi_F^3(G,\delta)$ is disconnected.
\end{proof}

\subsection{Efficiently recognizing graphs with connected and convex Cayley configuration spaces}

Theorems \ref{the:mainLooser} and \ref{the:2D3D} characterize
when a given graph always admits
(universally inherent)  connected,  convex, (generically 
complete),  linear polytope 2D and 3D (squared) Cayley configuration spaces. 
Based on these theorems, we give efficient algorithms 
to recognize these properties. 

\begin{theorem}
\label{the:algo1}
Given a graph $G$, 
both for the case of distance constraints as well
as for distance interval constraints,
there are linear time algorithms to 
recognize:
\begin{enumerate}
\item
whether
there is a non-empty set of non-edges $F$ such that $G$ 
always admits a connected 
(convex, linear polytope) 2D Cayley configuration
space on $F$. 
\item
under the assumption that $G$ is not over-constrained, whether there is a set of non-edges $F$ 
such that $G$ always admits a connected (convex, linear polytope) generically complete 2D Cayley configuration
space on $F$. 
\item
whether 
$G$ always admits  universally inherent, connected (resp. convex), 
$d$-dimensional (resp. squared) Cayley configuration
spaces, for $d\le 3$.
\end{enumerate}
\end{theorem}

\begin{proof}
For (1), first note that a linear time algorithm decomposes the 
input graph to 
2-sum components. By Theorem~\ref{the:mainLooser}, there is a non-empty 
set of non-edges $F$ such that $G$ 
always admits a connected 
(convex, linear polytope) 2D Cayley configuration
space on $F$, for the case of distance constraints if and only if $G$ has a 2-sum component that is
an underconstrained partial 2-tree. 
Thus, we only need to check whether there
exists a 2-sum component that is an underconstrained partial 2-tree.
Checking if a graph is a partial 2-tree can be done
in linear time. Since partial 2-trees are always independent, i.e, they 
cannot be overconstrained, 
they are underconstrained exactly if the number of edges is at most
twice the number of vertices minus 3 (the Laman count).
Thus the entire check can be done in linear time.

Under the assumption that $G$ does not have overconstrained subgraphs, the
algorithm for (2) is essentially the same as (1).
By Theorem \ref{the:new-oneEdgeProjection},
check whether all underconstrained 2-sum components are partial 2-trees.
Find any set of nonedges $F$ that complete them into 2-trees, which
are automatically wellconstrained or minimally rigid.
This guarantees a generically complete 
Cayley configuration space on $F$.
The assumption that $G$ has no overconstrained subgraphs is necessary 
for the above algorithm to work:
for example, suppose the input graph 
$G$ is a 2-sum of two graphs $G_1$ and $G_2$ where $G_1$ is an
underconstrained partial 2-tree while $G_2$ is not a partial 2-tree.
There is a complete set of non-edges  $F$ for $G_1$ such that
$G_1$ always admits a generically complete, 
connected, convex and linear polytope 2D Cayley configuration space. 
However, in order to check whether $F$ yields a generically complete
Cayley configuration space 
for all of $G$, we have to ensure that $G$ is rigid.
If no overconstrained subgraphs exist, then this can be determined
by a Laman count for $G_2$ in linear time. Otherwise,
we would need to detect the presence of overconstrained subgraphs.
In general,  $O(|E||V|)$ network flow based algorithms 
\cite{bib:HoLoSi01b, bib:survey}, 
and even more efficient 
pebble game algorithms 
\cite{bib:hendrickson} \cite{bib:streinu} 
exist for this check. 

For (1) and (2) in the case of distance interval constraints, Theorem \ref{the:interval}
permits us to perform the same checks and in linear time on a subdivision
$G'$ of $G$ that is at most 2 times the size of the edge set of $G$.

For (3), by Theorem~\ref{the:2D3D}, which applies to both distance
and distance interval constraints, we only need to check whether a graph
is 2-realizable( partial 2-tree) and 3-realizable. 
The former is straightforward. 
Based on \cite{bib:connelly05}, 
a linear algorithm is proposed in 
 \cite{bib:yinyuye06-2} to check
whether a graph is 3-realizable.

\end{proof}



\subsection{Practical use: Sampling and Realization Complexities}

Theorem~\ref{the:algo1} gives us the algorithms  to 
find a set of non-edges that ensure  (generically complete), connected,
convex, linear polytope Cayley configuration spaces.
They prove Points (1), (2), (3) of the Theorem below.
For practical use we need to further get the 
exact description of $\Phi_F^d(G,\delta)$ (resp. $\Phi_F^d(G, [\delta^l,\delta^r])$) (sampling complexity)
and also  a cartesian 
realization of a given parametrized configuration 
in $\Phi_F^d(G,\delta)$(resp. $(G, [\delta^l,\delta^r])$) (realization complexity).

\begin{theorem}
\label{the:algo2}
Given an EDCS $(G,\delta)$, or $(G,[\delta^l,\delta^r])$,
where the corresponding graphs $G$ 
are recognized by the algorithms of Theorem \ref{the:algo1},
there are linear time algorithms that:
\begin{enumerate}
\item
output 
a set of non-edges $F$ such that $G$ 
always admits a (generically complete) connected 
(convex, linear polytope) 2D Cayley configuration
space on $F$. 
\item
under assumption $G$ is not overconstrained, 
output
a set of  
non-edges $F$ such that $G$ 
always admits a connected 
(convex, linear polytope) generically complete 2D Cayley configuration
space on $F$.
\item
output 
a maximal set of non-edges $F$ such that the graph $H := G \cup F$
always admits  universally inherent, connected 2D or 3D Cayley configuration
spaces. 
\item
For $F$ in  all of the above items, output a corresponding 
description of a linear polytope 2D Cayley configuration
space $\Phi_F^2(G,\delta)$ (resp. 
$\Phi_F^2(G,[\delta^l,\delta^r])$) provided it is known to be nonempty; 
additionally, if one realization $p$ is known, then given any input 
element in the 2D Cayley configuration space, output its corresponding  realization
in $\mathbb{R}^2$. 
\end{enumerate}
Given an EDCS $(G,\delta)$, or $(G,[\delta^l,\delta^r])$,
there is a polynomial time algorithm that 
\begin{itemize}
\item[(5)]
for the $F$ in Item 3, outputs one configuration in the configuration
space $\Phi_F^3(G,\delta)$ (resp. 
$\Phi_F^3(G,[\delta^l,\delta^r])$); additionally, when given 
an input element
of this 3D Cayley configuration space, 
outputs a realization $\mathbb{R}^3$. 
\end{itemize}
\end{theorem}

\begin{proof}
As pointed out earlier, the proof of Theorem \ref{the:algo1}
also proves (1),(2). For (3), for 2D, we simply output a set of 
non-edges $F$ so that $G\cup F$ is a complete 2-tree.
For 3D, using a characterization theorem
of 3-realizable graphs $G$, by \cite{bib:connelly05, bib:slaugher04},  
and an algorithm of \cite{bib:yinyuye06-2}, 
we can output a maximal set of edges $F$ such that 
such that $G\cup F$ remains 3-realizable.
This gives the desired graph $H$ by Theorem \ref{the:EDMandRealizability}.

\medskip
For (4), for the case of 
distance constraints alone, the proof of Lemma \ref{lem:sufficient}
gives a linear time algorithm  to find the description of 
the linear polytope 2D configuration 
space on $F$ as a system of linear inequalities.
For the case of distance interval constraints, the same algorithm
can be applied to the subdivision graph $G'$ given by Theorem \ref{the:interval}
to get the description.
The assumption that the Cayley configuration space is non-empty is required
because by Lemma \ref{lem:2-sumProjection},
the Cayley configuration space of $G$
would be empty if any of EDCS corresponding to a non-partial-2-tree 2-sum components
of $G$ has no realization; and determining if a general EDCS has a 
realization is an NP-hard problem.

For the second statement in (4): take $G_C$ (resp. $G'_C$) to be any of 
the 2-sum components of $G$ (resp. $G'$) that contain a subset of $F_C \subseteq F$. 
When an element, $\delta^*(F_C)$ of  a 
Cayley configuration space of such an EDCS is given,  
if $G_C\cup F_C$ (resp. $G_C' \cup F_C$) 
are complete 2-trees,  the corresponding EDCS
$(G_C\cup F_C, \delta, \delta^*(F_C))$
(resp.  $(G'_C\cup F_C, \delta, \delta^*(F_C))$)
are Quadratically or Radically realizable. I.e., 
they can be realized by 
a straightforward
ruler-and-compass construction involving a sequence of solutions
 of  univariate quadratics (intersection of circles). 
On the other hand, if  $G_C\cup F_C$ (resp. $G_C' \cup F_C$) 
are not complete 2-trees, they can be made complete 2-trees 
by additional edges $D_C$, for which $\delta$ can be extended
consistently, i.e., in such a way that 
$(G_C\cup D_C \cup F_C, \delta, \delta(D_C), \delta^*(F_C))$
has a realization whenever 
$(G_C \cup F_C, \delta,  \delta^*(F_C))$
has a realization.
The realizations of the remaining 
2-sum components (that do not contain any subset of $F$), 
in particular those that may not be partial 2-trees,
can be read off from the one given realization $p$. 
As shown in Lemma \ref{lem:2-sum},
realizations of the different 2-sum components do not 
interfere with each other: i.e., the realizations of the 
various 2-sum components can be 
hinged together along the 2-sum edges to get the realization
of the entire graph.
Again, the realization $p$ is required because finding a realization for 
the EDCS of a 2-sum component that is a general graph (not necessarily
a partial 2-tree) is a hard problem (the decision version is NP-hard
\cite{bib:saks}). 

\medskip
For (5), 
simply using the definition of 3-realizable graphs, we can obtain 
one point $\delta^*(F)$ of the Cayley configuration space of $G$ on the given $F$  
as follows. Use the positive semidefiniteness of the Euclidean
distance matrix cone to complete the distances
or distance intervals (for the edges of $G$ as given by $\delta$) 
into a full Euclidean distance matrix
for some dimension $d$.  This $d$ could be much larger than 3.
Since we know that the given $F$ satisfies the property that
$H = G \cup F$ is also 3-realizable, 
by the definition of 3-realizability, if we pick the 
distances for $\delta^*(F)$ from the completed matrix, 
then it would be guaranteed to be a point in 
the Cayley configuration space of $G$ on $F$.

Now given such a $\delta^*(F)$ for a such a maximal set $F$,  
using a characterization of \cite{bib:connelly05,bib:slaugher04}  
of 3-realizable graphs as 2-sums and 3-sums of a small number
of special types of graphs, the paper of \cite{bib:yinyuye06-2}
gives an algorithm to get one realization 
of $(G\cup F, \delta, \delta^*(F))$ in $\mathbb{R}^3$.
\end{proof}

%% file: conclusions.tex
\section{Conclusions and Conjectures}
\label{sec:conclusion}
Our results give a practically meaningful, and mathematically robust
definition of exact and efficient Cayley configuration spaces 
of underconstrained  Euclidean Distance Constraint Systems (equalities
and inequalities), based on 
various efficiency factors including
complexity of sampling and realization.
We have taken the first step in a systematic and graded program
sketched in \cite{bib:GaoSitharam05} and laid out in \cite{bib:Gao08} - for the
combinatorial characterizations of efficient Cayley configuration spaces.
In particular, the results presented here characterize
graphs and their Cayley parameters that yield  (squared)
Cayley configuration spaces that are connected, convex, linear polytopes,
and efficient algorithms for sampling realizations. 

\subsection{Theoretical Directions and Conjectures}

\subsubsection{Is $d$-realizability equivalent to 
universally inherent,
connected $d$-dimensional Cayley configuration spaces?}
\label{subsec:conjecture1} 

Our first conjecture is the reverse direction of Theorem \ref{the:EDMandRealizability}.

\begin{conjecture}
For any dimension $d$,  a graph is $d$-realizable 
if and only if it always admits universally inherent, connected
(resp. convex) $d$-dimensional squared Cayley configuration spaces.
\end{conjecture}
\label{sec:conjeture1}

We can try to leverage
results \cite{bib:connelly, bib:Gortler} about connected components of the $d$-dimensional 
realization spaces
of a graph based on its higher, $d'$-dimensional realization spaces,
where $d' > d$.

\subsubsection{The roles of Genericity and Independence}
\label{sec:globally-linked}

\begin{figure}
\psfrag{v1}{$v_1$}
\psfrag{v2}{$v_2$}
\psfrag{v3}{$v_3$}
\psfrag{v4}{$v_4$}
\psfrag{v5}{$v_5$}
\psfrag{f}{$f$}
\centerline{ 
\includegraphics[width=5cm]{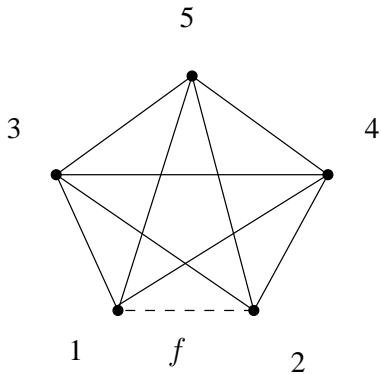}
}
\caption{  
A generically globally rigid graph in 2D.
}
\label{F:globally-linked}
\end{figure}

Our results characterize (for the case of distance equalities and frameworks) 
Cayley configuration space properties that hold for {\sl all}
distance assignments $\delta$, 
and are hence incorrect if we require the properties
to hold only in generic situations.
For example, Figure \ref{F:globally-linked}
shows a generically globally rigid graph $G$: the generic
2D Cayley configuration space of this graph on the non-edge $f$ is a single
point. However, by our Theorem \ref{the:new-oneEdgeProjection}, since a minimal 
2-sum component containing $f$ is not 2-realizable,
the Cayley configuration space on $f$ is disconnected.
The apparent discrepancy arises because the proof of the Lemma \ref{the:new-graph2} uses
non-generic specializations of the edge distances $\delta$ in
the process of reduction to a minor which then shows that the
Cayley configuration space on $f$ is disconnected {\sl for that $\delta$},
which is sufficient to prove the statement of
Lemma \ref{the:new-graph2} and Theorem \ref{the:new-oneEdgeProjection}.
We believe, however that both these results still hold under a genericity
assumption, {\sl provided} the graph $G$ is not overconstrained
(independent, in rigidity terminology) 
and the non-edge $f$ is not globally-linked.

\begin{conjecture}
Given a  graph $G$ that is not generically over-constrained 
and a non-edge $f$, the Cayley configuration space 
$\Phi_f^2(G,\delta)$ is a single interval 
for all {\sl generic} $\delta$ (i.e., for $\delta$ that admit a 
2D generic realization of $(G,\delta)$), 
if and only if all the minimal 2-sum components containing  
$f$ are 2-realizable and (partial  2-trees).
\end{conjecture}

Proving this would formally establish the connections
between our work and the work on generic global rigidity 
and generic globally linked pairs, 
including results of \cite{bib:connelly, bib:Gortler,bib:hendrickson-uniquerel, bib:jackson05} and
\cite{bib:jackson06}.

\subsubsection{Sampling complexity for 3D}

Note that while we know from Theorem \ref{the:2D3D} that 
3-realizable graphs always admit universally inherent, 
convex, 3D squared Cayley configuration spaces, we do not
yet  know an efficient algorithm to determine their 
description. This is necessary to determine the sampling complexity. 
  This is in stark contrast to the linear time
algorithm for obtaining such descriptions in the case of 2D
(Theorem \ref{the:new-EdgesetProjection}).

One straightforward algorithm for obtaining the 
description of  the Cayley configuration space 
$\Phi_F^d(G,\delta)$ as a semi-algebraic
set is to start with the Cayley-Menger  \cite{bib:cayley, bib:blumenthal, bib:menger}
determinantal equalities and inequalities 
for Euclidean distance matrices in $d$-dimensions.
These are polynomial relationships 
between all the ${|V|\choose 2}$ Cayley parameters,
including  those in $E$, $F$ and those in 
$\overline{E\cup F}$.
Eliminating those in 
$\overline{E\cup F}$, and specifying the values for the parameters in 
in $E$ to be $\delta(E)$ leaves polynomial 
equalities and inequalities in the Cayley parameters in $F$.
This yields the required 
description of  the Cayley configuration space 
$\Phi_F^d(G,\delta)$ as a semi-algebraic
set.

Viewed in this manner, it appears 
remarkable that in 2D, for the graphs $G$ and non-edge sets $F$ 
satisfying the conditions of Theorem \ref{the:new-EdgesetProjection}, this above-described elimination
leaves only the triangle inequalities relating 
the Cayley parameters in $E$ and $F$ (note that these were part
of the original Cayley-Menger set of inequalities) and hence we
get a linear polytope description of 
$\Phi_F^2(G,\delta)$. 
Note however that {\sl we did not use such an elimination for our proof!}. 
Our proof that these triangle inequalities 
give a description of  
$\Phi_F^2(G,\delta)$ 
was through a more direct route: 
we determined for what configuration $\delta^*(F)$ 
for the Cayley parameters in $F$
we could construct a 2D realization for the augmented EDCS
$\Phi_F^2(G\cup F,\delta(E),\delta(F^*)$. 

We would like to show a similar result in 3D either by using
elimination or using Euclidean distance matrix completion for fixed
rank \cite{bib:alfakih99, bib:yinyuye06}
by a more direct route of determining when
3D realizations  can be constructed.

\begin{conjecture}
Let $H$ be  a 3-realizable graph on vertex set $V$. 
Take any partition of the edge set of $H$ 
into $E\cup F$, and consider the graph $G = (V,E)$
and any EDCS $(G, \delta)$. 
Then there is a $O(|V|^2)$ time algorithm to 
write down the description of  the Cayley configuration space 
$\Phi_F^2(G,\delta)$ as a semi-algebraic 
set of low degree (say, no more than 4).
\end{conjecture}

In this context, note that there is also room to improve the sampling
complexity and obtain 
more refined descriptions 
even for 2D Cayley configuration spaces of 
graphs satisfying the conditions of Theorem \ref{the:new-EdgesetProjection}.
For this purpose, it makes sense to study
complete 2-trees $H$ on a set of vertices $V$.
In particular, after partitioning the edges of $H$ into
$E \cup F$ and 
eliminating all except the set of Cayley parameters in $E$
we obtain a system of linear inequalities in those parameters,
which we call the {\em 2D admissible distance} polytope of 
$G  = (V,E)$. I.e, it specifies the distance assignments $\delta$
for which $(G,\delta)$ has a 2D realization. 
Note that this is the projection of the 2D admissible distance polytope 
for $H$, onto the parameters in $E$. 
Furthermore, the Cayley configuration space 
$\Phi_F^2(G,\delta)$ is a {\sl section or a fibre} of this polytope.
In \cite{bib:BorceaSitharamStreinu} we study the detailed topology of these
polytopes, towards extending
similar studies of polytopes corresponding to the
simpler class of polygonal linkages
parametrized by Cayley parameters, see  for example
\cite{bib:hausman}.

\subsubsection{Characterizing all parameters that 
ensure connected, Cayley squared configuration spaces for 3D}
\label{sec:3dccs}

It would be desirable to extend Theorem \ref{the:new-oneEdgeProjection} (and hence 
Theorems \ref{the:mainLooser} and \ref{the:new-EdgesetProjection})
to a characterization of parameter choices for 
3D connected Cayley configuration spaces. 
As pointed out in Observation \ref{obs:3dCounter} the obvious analog
of this result fails in 3D.
Partial results supporting the following conjecture 
have been reported in \cite{bib:Gao08}.
Note that partial 3-trees are a large proper subclass 
of 3-realizable graphs. 

\begin{conjecture}
\label{con:partial3-Tree}
Given graph $G$ that is a partial 3-tree and non-edge $f$

\begin{itemize}
\item
if $G \cup f$ has no $K_{5}$ or
$K_{2,2,2}$ minor, then $G$ has a connected 3D Cayley configuration space on
$f$; 

\item 
if $G \cup f$ has a $K_{5}$ or
$K_{2,2,2}$ minor then 
$G$ has a connected 3D Cayley configuration space on
$f$ if and only if
the 2 vertices of $f$ must be identified in order to get
a $K_{5}$ or
$K_{2,2,2}$ minor in $G$. 
\end{itemize}
\end{conjecture}

\subsubsection{1-dof 2D mechanisms and 
2D Cayley configuration spaces with 2 connected components}

There are 2 possible directions 
to move beyond the connected, convex, linear-polytope 2D
configuration spaces. 

The first is to study the Cayley configuration spaces of 
1-dof graphs that {\sl just fail} to satisfy the 
requirements of Theorem \ref{the:new-graph2}. A first step in this direction has been 
taken by \cite{bib:GaoSitharam08b}, see also \cite{bib:Gao08}, and
future research suggestions along that direction has been presented 
there.

The second direction is to study which graphs always admit 
Cayley configuration spaces with 2 or fewer connected components?
Partial results in this direction 
have been reported in \cite{bib:Gao08},
specifically towards 
characterizing graphs $G$ and non-edges $f$ 
such that the 2D Cayley configuration space of $G$ on $f$ has no more
than 2 intervals.

\subsubsection{Stronger results for special distance assignments}
\label{sec:special-distance}
Our characterizations apply to all distance assignments $\delta$ or 
distance intervals $[\delta^l,\delta^r]$.
This places strong restrictions on the characterized classes of graphs $G$ with
efficient Cayley configuration spaces. 
However, if one assumes 
special distance assignments, well-behaved Cayley configuration spaces
may exist for much larger classes of graphs. For example, consider the 
2D 2-direction grid, which is not a partial 2-tree.  
However, under the restriction of unit-distance edges,
\cite{bib:Gao08} shows that such an EDCS has a Cayley configuration space
that is a convex polytope, in fact a {\em rectilinear box}.
Moreover, for such an EDCS,
\cite{bib:Gao08} gives 
a complete characterization of Cayley parameters (non-edges)
on which the Cayley configuration space is convex.
It would be desirable to obtain similar results for 3D,
due to potential applications in Section \ref{sec:practical} below.

\subsection{Practical Implications for 
Mechanical CAD and Computational Chemistry}
\label{sec:practical}

\subsubsection{Realization space and motion exploration for CAD mechanisms}

Representation of realization spaces of underconstrained 
systems is a long-standing issue in the development of 
constraint solvers underlying mechanical CAD systems.
Our characterizations and algorithms for obtaining efficient configuration
spaces and sampling realizations  would be  useful to 
incorporate into commercial constraint solvers.
However, they need to be implemented in combination with other practical
user- or designer-driven functionalities.

In addition to sampling realizations, 
our program of comprehensive study of 
possible parameter choices
is useful for {\sl geometrically meaningful exploration}
of the motions of underconstrained systems, or mechanisms.
Current (infinitesimal) motion representation reduces to 
a basis choice for the motion space obtained from the rigidity
matrix for a particular realization $p$.
Obvious basis choices lead to an erratic exploration of the motion space,
often amplifying spurious ``globally" coupled motions or ``allosteric" effects 
instead of systematically 
first exploring ``locally" coupled motions to the
extreme or boundary configurations.

Intuitively, 
our theorems give choices of Cayley parameters that ensure convex or otherwise
well-behaved Cayley configuration spaces and hence each standard method
of walking these spaces invests a geometric meaning 
to the corresponding systematic motion exploration, 
especially since Cayley parameters are particularly suited to representing 
internal motion.
Furthermore
any standard method of walking a convex Cayley configuration space 
would fully explore ``locally"  coupled motions before systematically
progressing to less local ones.
It would be desirable to state and prove a formal statement to this effect.

\subsubsection{Sampling realization space for helix packing}
\label{sec:helix}
Helix packing is a well-studied computational chemistry problem
since helices are prevalent in many biomolecules.
One part of the problem can be viewed as sampling the realization space of an 
EDCS (including distance inequalities) that correspond to 
steric or collision avoidance and other constraints. 
Current methods use inexact representations of realization
spaces and ``generate and test" algorithms for sampling them. As a result
they are inefficient and moreover lack any guarantee that all boundary
and extreme realizations have been explored.
 
Partial results \cite{bib:Gao08} (see Section \ref{sec:3dccs}) 
give parameter choices that ensure connected 
and convex, 3D squared Cayley configuration spaces 
{\it even for  non 3-realizable graphs}. By the characterization proved here, 
these are clearly not universally inherent. Based on these
Cayley parameter choices, an algorithm for 
exact description
and efficient, systematic sampling of the helix packing configuration 
spaces (especially the boundaries and extreme points) is given in
\cite{bib:helix}. 

\subsubsection{Limit properties and Sampling of 
realization spaces  for Zeolite sequences} 
\label{sec:zeolite}
Zeolites are naturally occuring materials whose structure can be
described as ``corner sharing regular tetrahedra."
The corresponding EDCS sequences involve only distance equalities and 
are obtained from progressively larger pieces of an 
infinite lattice structure with uniform boundary conditions. 
The goal is to sample the realization spaces of these Zeolite
EDCS sequences, with a view to understanding their properties
in the limit.

Since the Zeolite EDCS have special distance assignments, 
as pointed out in Section \ref{sec:special-distance},
they have efficient Cayley configuration spaces 
even if the corresponding graphs may not satisfy
the requirements of the characterization theorems presented 
in this manuscript. Using similar 
results such as a characterization of Cayley configuration spaces
of 2D unit-distance grid graphs mentioned in Section \ref{sec:special-distance},
\cite{bib:Gao08} presents partial results on the configuration
spaces of a class of Zeolite sequences.

subsection*{Acknowledgements}
Special thanks to the participants of the 
Barbados (McGill) workshops on Rigidity and molecular modeling 
organized by Ileana Streinu and Mike Thorpe
in 2006 and 2007. Those workshops helped in discovering 
several of the applications listed in Section \ref{sec:practical}.